\RequirePackage{etoolbox}
\csdef{input@path}{%
 {sty/}
}%

\documentclass[ba]{imsart}
\pubyear{0000}
\volume{00}
\issue{0}
\doi{0000}
\firstpage{1}
\lastpage{1}

\usepackage{amsthm}
\usepackage{amsmath}
\usepackage{natbib}
\usepackage[colorlinks,citecolor=blue,urlcolor=blue,filecolor=blue,backref=page]{hyperref}
\usepackage{graphicx}
\usepackage{extarrows}

\startlocaldefs
\numberwithin{equation}{section}
\theoremstyle{plain}

\endlocaldefs

\begin{document}

 \begin{frontmatter}
\title{Using stacking to average Bayesian predictive distributions}
\runtitle{Using stacking to average Bayesian predictive distributions}

\begin{aug}
\author{\fnms{Yuling} \snm{Yao}\thanksref{addr1}\ead[label=e1]{yy2619@@columbia.edu}},
\author{\fnms{Aki} \snm{Vehtari}\thanksref{addr2}\ead[label=e2]{ Aki.Vehtari@aalto.fi}}
\author{\fnms{Daniel} \snm{Simpson}\thanksref{addr3}\ead[label=e3]{dp.simpson@gmail.com}}
\and
\author{\fnms{Andrew} \snm{Gelman}\thanksref{addr4}\ead[label=e4]{gelman@stat.columbia.edu}}
\runauthor{Y. Yao, A. Vehtari,  D. Simpson and A. Gelman}
\address[addr1]{Department of Statistics, Columbia University, New York, NY,  \printead{e1} }
\address[addr2]{Helsinki Institute of Information Technology, Department of Computer Science, Aalto University,  Finland,  \printead{e2} }
\address[addr3]{Department of Statistical Sciences, University of Toronto, Canada,  \printead{e3} }
\address[addr4]{Department of Statistics and Department of Political Science, Columbia University, New York, NY,  \printead{e4} }
\end{aug}

\begin{abstract} 
The widely recommended procedure of Bayesian model averaging is flawed in the $\mathcal{M}$-open setting in which the true data-generating process is not one of the candidate models being fit. We take the idea of {\em stacking} from the point estimation literature and generalize to the combination of predictive distributions, extending the utility function to any proper scoring rule, using Pareto smoothed importance sampling to efficiently compute the required leave-one-out posterior distributions and regularization to get more stability. We compare stacking of predictive distributions to several alternatives:  stacking of means, Bayesian model averaging (BMA), pseudo-BMA using AIC-type weighting, and a variant of pseudo-BMA that is stabilized using the Bayesian bootstrap. Based on simulations and real-data applications, we recommend stacking of predictive distributions, with BB-pseudo-BMA as an approximate alternative when computation cost is an issue.
\end{abstract}

\end{frontmatter}

\section{Introduction}

A general challenge in statistics is prediction in the presence of multiple candidate models or learning algorithms $\mathcal{M}=(M_1,\dots, M_K)$. Model selection---picking one model that can give optimal performance for future data---can be unstable and wasteful of information \citep[see, e.g.,][]{predictiveCompare}.  An alternative is model averaging, which tries to find an optimal model combination in the space spanned by all individual models.   In Bayesian context, the natural target for prediction is to find a predictive distribution that is close to the true data generating distribution \citep{score,Vehtari+Ojanen:2012}.   
  
Ideally, we prefer to attack the Bayesian model combination problem via continuous model expansion---forming a larger bridging model that includes the separate models $M_k$ as special cases \citep{gelman2011parameterization}---but in practice constructing such an expansion can require conceptual and computational effort, and so it makes sense to consider simpler tools that work with existing inferences from separately-fit models.

\subsection{Bayesian model averaging}
If the set of candidate models between them represents a full generative model, then the Bayesian solution is to simply average the separate models, weighing each by its marginal posterior probability.  This is called {\em Bayesian model averaging} (BMA) and is optimal if the prior is correct, that is, if the method is evaluated based on its frequency properties evaluated over the joint prior distribution of the models and their internal parameters \citep{madigan1996bayesian, hoeting1999}.  If $y=(y_1, \dots,y_n)$ represents the observed data, then the posterior distribution for any quantity of interest $\Delta$ is, 
$$p(\Delta | y)= \sum_{k=1}^K p(\Delta| M_k, y)p(M_k |y).$$
In this expression, each model is weighted by its posterior probability,
$$p(M_k | y)  =\frac{p(y | M_k ) p(M_k)} {\sum_{k=1}^K p(y | M_k ) p(M_k)} ,$$
and this expression depends crucially on the marginal likelihood under each model,
$$p(y | M_k ) = \int\! p(y | \theta_k, M_k ) p(\theta_k | M_k)  d \theta_k\,.$$

In Bayesian model comparison, the relationship between the true data generator and the  model list $\mathcal{M}=(M_1,\dots, M_K)$ can be classified into three categories: $ \mathcal{M}$-closed, $ \mathcal{M}$-complete  and  $ \mathcal{M}$-open. We adopt the following definition from  \citet{Bernardo+Smith:1994, key1999bayesian, clyde2013bayesian}: 
\begin{itemize}
\item $ \mathcal{M}$-closed means the true data generating model is one of $M_k \in \mathcal{M}$, although it is unknown to researchers.  
\item $\mathcal{M}$-complete refers to the situation where the true model exists and is out of model list $\mathcal{M}$. But we still wish to use the models in $\mathcal{M}$ because of  tractability of computations or communication of results, compared with the actual belief model.  Thus, one simply finds the member in $\mathcal{M}$ that maximizes the expected utility (with respect to the true model).
\item $\mathcal{M}$-open  refers to the situation in which we know the  true model $M_t$ is not in $ \mathcal{M}$, but we cannot specify the explicit form $p(\tilde y|y) = p(\tilde y|M_t,y)$ because it is too difficult, we lack time to do so, or do not have the expertise, computational intractability, etc. 
\end{itemize}

BMA is appropriate  for $\mathcal{M}$-closed case.  In the $\mathcal{M}$-open and $\mathcal{M}$-complete case,  BMA will asymptotically  select  the one  single model in the list that is closest in Kullback-Leibler (KL) divergence.

Furthermore, in BMA, the marginal likelihood depends sensitively on the specified prior $p(\theta_k | M_k)$ for each model. For example, consider a problem where a parameter has been assigned a normal prior distribution with center 0 and scale 10, and where its estimate is likely to be in the range $(-1, 1)$. The chosen prior is then essentially flat, as would also be the case if the scale were increased to 100 or 1000. But such a change would divide the posterior probability of the model by roughly a factor of 10 or 100.

\subsection{Predictive accuracy}
From another direction, one can simply aim to minimize out-of-sample predictive error, equivalently to maximize expected predictive accuracy. In this paper we propose a novel log score stacking method for combining Bayesian 
predictive distributions. As a side result we also propose a simple model weighting using Bayesian leave-one-out cross-validation.

\subsection{Stacking}
{\em Stacking} is a direct approach for averaging point estimates from multiple models.  The idea originates with \cite{wolpert1992}, and \cite{breiman1996} gives more details for stacking weights under different conditions.  In supervised learning where the data are $((x_i, y_i), i=1,\dots, n)$, and each model $M_k$ has a parametric form $\hat y_k=f_k(x | \theta_k)$, stacking is done in two steps \citep{ting1999}. In the first, baseline-level, step, each model is fitted separately and the leave-one-out (LOO) predictor $\hat f_k ^{(-i)}(x_i) =E(y_i | \hat \theta_{k ,y_{-i}}, M_k )$ is obtained for each model $k$ and each data point $i$. Cross-validation or bootstrapping is used to avoid overfitting \citep{leblanc1996}.   In the second, meta-level, step, a weight for each model is obtained by minimizing the mean squared error, treating the leave-one-out predictors from the previous stage as  covariates:
\begin{equation}\label{stackingMean}
 \hat w=\arg \min_w \sum_{i=1}^n \left(y_i-\sum_k w_k \hat f_k^{(-i)}(x_i)\right) ^2. 
 \end{equation} 
\cite{breiman1996} notes that a positive constraint $w_k \geq  0,  k=1,\dots K$, or a $K-1$ simplex constraint: $w_k \geq  0,  \sum_k^{K} w_k =1$ enforces a solution.  Better predictions may be attainable using regularization \citep{merz1999,yang2014minimax}.  Finally,   the  point prediction for a new data point with feature vector $\tilde x $ is $$ \hat {\tilde y}= \sum_{k=1}^K \hat w_k f_k(\tilde x |  \hat \theta_{k ,y_{1:n}}  ).$$

It is  not surprising that  stacking typically outperforms  BMA when the criterion is mean squared predictive error  \citep{clarke2003}, because BMA is not optimized to this task.  \citet{wong2004improvement} emphasize that the BMA weights reflect the fit to the data rather than evaluating the prediction accuracy.  On the other hand, stacking is not widely used in Bayesian model combination because it only works with point estimates, not the entire posterior distribution \citep{hoeting1999}.

\citet{clyde2013bayesian} give a Bayesian interpretation for stacking  by considering  model combination  as a decision problem when  the true model $M_t$ is not in the model list.  If the decision is of the form $a(y,w)=\sum_{k=1}^K w_k \hat y_k $, then the expected utility under quadratic loss is,
$$\mathrm{E}_ {\tilde y} (u( \tilde y, a(y,w) )| y ) =- \int || \tilde y- \sum_{k=1}^K w_k \hat {\tilde {y}}_k ||^2 p(\tilde y | y, M_t ) d\tilde y, $$
where $\hat {\tilde {y}}_k$ is the predictor of new data $\tilde y$ in model k.
The stacking weights are the solution to the LOO estimator:
$$\hat w= {\arg\max} _w \frac{1}{n}  \sum_{i=1}^n u(y_i, a(y_{-i}, w)   ) ,$$
where $a(y_{-i}, w)=\sum_{k=1}^K w_k \mathrm{E}(y_i | y_{-i}, M_k)$.

 The stacking predictor for new data $\tilde y$ is  $ \sum_{k=1}^K \hat w_k \hat {\tilde {y}}_k$. The predictor in the $k$-th model $\hat {\tilde {y}}_k$ can be  either the plug-in  estimator ,
 $$ \mathrm{E}_k(\tilde  y  | \hat \theta_{k}, y )= \int \tilde  y  p (\tilde  y  |\hat \theta_{k, y}, M_k )  d\tilde  y, $$
  or the posterior  mean,
  $$\mathrm{E}_k(\tilde  y  | y ) = \int \tilde y p( \tilde  y | \theta_{k} , M_k)  p(\theta_k| y, M_k)  d \tilde y d \theta_k. $$

 \cite{le2016bayes} prove that the stacking solution is asymptotically the Bayes solution.  With some mild conditions on the distribution, the following asymptotic  relation holds:
$$  \int l( \tilde  y, a(y,w) ) p(\tilde  y | y)   d \tilde  y    - \frac{1}{n}\sum_{i=1}^n l( y_i, a(y_{-i}, w) )  \xlongrightarrow{\text{$L_2$}} 0, $$
where $l$ is the squared loss, $l(\tilde y,a)=(\tilde y-a)^2.$
They also prove when the action is a predictive distribution  $a(y_{-i}, w)=\sum_{k=1}^K w_k p(y_i | y_{-i}, M_k)$, then the  asymptotic  relation still holds for negative logarithm scoring rules. 

However,  most early literature limited  stacking to  averaging \emph{point} predictions, rather than  \emph{predictive distributions}. In this paper, we extend stacking from minimizing the squared error to maximizing scoring rules,  hence make stacking applicable to combining a set of Bayesian posterior predictive distribution.   We argue this is the appropriate version of Bayesian model averaging in the $\mathcal{M}$-open situation. 

\subsection{Fast leave-one-out cross-validation}
Exact leave-one-out cross validation can be computationally costly.  For example, in the econometric literature,  \citet{geweke2011optimal,geweke2012prediction} suggest averaging prediction models by maximizing predictive log score, only considering time series due to the computational cost of exact LOO for general data structures.  In the present paper we demonstrate that Pareto-smoothed importance sampling leave-one-out cross-validation (PSIS-LOO) \citep{practicalPSIS} is a practically efficient way to calculate the needed leave-one-out predictive densities $p(y_i | y_{-i}, M_k)$ to compute log score stacking weights.

\subsection{Akaike weights and pseudo Bayesian model averaging}

Leave-one-out cross-validation is related to various information criteria \citep[see, e.g.][]{Vehtari+Ojanen:2012}. In case of maximum likelihood estimates, leave-one-out cross-validation is asymptotically equal to Akaike's information criterion (AIC) \citet{Stone:1977a}. 
Given AIC =-2 log (maximum likelihood) + 2 (number of parameters), \citet{akaike1978likelihood} proposed to use $\exp(-\frac{1}{2} \mathrm{AIC})$ for model weighting \citep[see also][]{Burnham+Anderson:2002,wagenmakers2004aic}.
More recently we have seen also Watanabe-Akaike information criterion (WAIC) \citep{Watanabe:2010d} and leave-one-out cross-validation estimates used to compute model weights following the idea of AIC weights.

In Bayesian setting \citeauthor{Geisser+Eddy:1979} (\citeyear{Geisser+Eddy:1979}; see also, \citealt{Gelfand:1996}) proposed pseudo Bayes factors where marginal likelihoods $p(y|M_k)$ are replaced with a product of Bayesian leave-one-out cross-validation predictive densities $\prod_{i=1}^np(y_i | y_{-i}, M_k)$. Following the naming by \citeauthor{Geisser+Eddy:1979}, we call AIC type weighting which uses Bayesian cross-validation predictive densities as pseudo Bayesian model averaging (Pseudo-BMA).

In this paper we show that the uncertainty in the future data distribution should be taken into account when computing such weights. We will propose a AIC type weighting using the Bayesian bootstrap and the expected log predictive density (elpd), which we call Pseudo-BMA+ weighting. We show that although Pseudo-BMA+ weighting gives better results than regular BMA or Pseudo-BMA weighting (in $\mathcal{M}$-open setting), it is still inferior to the log score stacking. Due to its simplicity we use Pseudo-BMA+ weights as initial guess for optimization procedure in the log score stacking.

\subsection{Other model weighting approaches}\label{reference-bma}

Besides BMA, stacking and AIC type weighting, some other methods have been introduced to combine Bayesian models.    \citet{gutierrez2005statistical}  consider using a nonparametric prior in the  decision problem stated above.  Essentially they are fitting a mixture model with a Dirichlet  process, yielding a posterior expected utility of,
$$  U_n (w_k, \theta_k)= \sum _{i=1}^{n}  \sum_{ k=1}^K w_k f_k (y_i | \theta_k).$$
They then solve for the optimal  weights $\hat w_k = \arg \max_{w_k, \theta_k}U_n (w_k, \theta_k)$.

\citet{dbd} propose model averaging using weights based on divergences from a reference model in $\mathcal{M}$-complete setting. If the true data generating density function  is known to be  $f^*$, then an AIC type (or Boltzmann-Gibbs type) weight can be defined as,
\begin{equation} \label{reference-pseudo-bma}
  w_k =\frac{\exp\bigl(-n \mathrm{KL}(f^*, f_k) \bigr)}{\sum_{k=1}^K  \exp \bigl(-n \mathrm{KL}(f^*, f_{k})\bigr)  }.
\end{equation}
The true model can be approximated with a reference model $M_0$ with density  $f_0( . | \theta_0)$ using nonparametric methods like Gaussian process or Dirichlet process, and $\mathrm{KL}(f^*, f_k)$ can be estimated by its posterior mean,
$$ \mathrm{\widetilde{ KL}_1} (f_0, f_k)=\int \!\!\int \mathrm{KL}  \Bigl( f_0(\cdot| \theta_0) , f_k(\cdot| \theta_k) \Bigr)p(\theta_k | y, M_k)p(\theta_0 | y,  M_0) d\theta_k d\theta_0,$$ 
or by the Kullback-Leibler divergence for posterior predictive distributions,
$$ \mathrm{\widetilde{ KL}_2} (f_0, f_k)= \mathrm{KL} \Bigl( \int\! f_0(\cdot| \theta_0)p(\theta_0 | y, M_0)d\theta_0 , \int\! f_k(\cdot| \theta_k) )p(\theta_k | y, M_k)d\theta_k  \Bigr) .$$ 
Here, $\mathrm{\widetilde{ KL}_1}$ corresponds to Gibbs utility, which can be criticized for not using the posterior predictive distributions \citep{Vehtari+Ojanen:2012}, although asymptotically the two utilities are identical, and $\mathrm{\widetilde{ KL}_1}$ is often computationally simpler than $\mathrm{\widetilde{ KL}_2}$.

Let $p(\tilde y |y, M_k)  = \int\! f_k(\tilde y| \theta_k) p(\theta_k | y, M_k)d\theta_k $,  $k=0,\dots,K $, then 
$$\mathrm{\widetilde{ KL}_2}  (f_0, f_k)= -\int\! \log p(\tilde y |y, M_k) p(\tilde y |y, M_0) d \tilde y +  \int\! \log p(\tilde y |y, M_0) p(\tilde y |y, M_0) d \tilde y $$ 
As the entropy of the reference model $\int\! \log p(\tilde y |y, M_0) p(\tilde y |y, M_0) d \tilde y$ is constant, the corresponding terms cancel out in the weight (\ref{reference-pseudo-bma}), leaving
$$w_k = \frac{  \exp \bigl(n \int\! \log p(\tilde y |y,M_k) p(\tilde y |y, M_0) d \tilde y  \bigr) }  {\sum_{k=1}^K   \exp \bigl(n \int\! \log p(\tilde y |y, M_{k}) p(\tilde y |y, M_0) d \tilde y   \bigr)}$$
It is proportional to the exponential expected log predictive density, where the expectation  is taken with respect to the reference model $M_0$. Comparing with  definition \ref{P-BMA} in Section \ref{Pseudo-BMA}, this method could be called Reference-Pseudo-BMA.

\section {Theory and methods}

We label  classical  stacking (\ref{stackingMean}) as \emph{stacking of means} because it  combines the models by minimizing the mean squared error of the point estimate, which is the $L_2$ distance between the posterior mean and  observed data.
In general, we can use a  proper scoring rule (or equivalently  the underlying divergence) to compare distributions.   After choosing that,  stacking can be  extended to combining  the whole distribution. 

\subsection{Stacking using proper scoring rules}
Adapting the notation of  \cite{score}, we label $Y$ as the random variable on the sample space $(\Omega, \mathcal{A})$ that can take values on $(-\infty, \infty)$. $\mathcal{P}$ is a convex class of probability measure on $\Omega$.  Any member of $\mathcal{P}$ is called a probabilistic forecast. 
A function $S:  \mathcal{P} \times \Omega \to  \bar{ \mathrm{R}}$   defines a scoring rule if $S(P, \cdot)$ is  $\mathcal{P}$ quasi-integrable for all  $P \in \mathcal{P}$.  In the continuous case,  distribution $P$ is identified with density function $p$.

For two probabilistic forecasts $P$ and $Q$,  we write $S(P, Q)=\int S(P, \omega) dQ(\omega) $.  A scoring rule $S$ is called \emph{proper} if $S(Q,Q) \geq S(P,Q)$  and \emph{strictly proper} if the equation holds only when $P= Q $ almost surely.   A proper scoring rule defines the divergence $d: \mathcal{P} \times \mathcal{P} \to  (0, \infty) $ as  $d(P,Q)=S(Q,Q)-S(P,Q)$. 
For continuous variables, some popularly  used scoring rules include:
\begin{enumerate}
\item {\em Quadratic score:}  $\mathrm{QS}(p,y) = 2p(y)-||p||_2^2$ with the divergence $d(p,q)= ||p-q ||_2^2$.

\item {\em Logarithmic score:}
$\mathrm{LogS}(p,y)=\log(p(y)) $ with $d(p,q)=\mathrm{KL}(q,p).$
The logarithmic score is the only proper local score assuming regularity conditions. 
 
\item {\em Continuous  ranked  score:}
$\mathrm{CRPS}(F,y)=-\int_\mathrm{R} (F(y')  - 1(y' \geq y) ) ^2 dy'$ with  $d(F, G)=\int_\mathrm{R} (F(y)-G(y))^2 dy $, where $F$ and $G$ are the corresponding distribution function.

\item {\em Energy score:}
$\mathrm{ES}(P,y)= \frac{1}{2}\mathrm{E}_P || Y-Y' ||_2^\beta-  \mathrm{E}_P|| Y-y||_2^\beta $, where $Y$ and $Y'$ are two independent random variables  from distribution $P$.
When $\beta=2$, this becomes $\mathrm{ES}(P,y)= -||\mathrm{E}_P(Y)-y||^2.$
The energy score is strictly proper when $\beta \in (0,2)$ but not when $\beta=2$.  

\item {\em Scoring rules depending on first and second moments:}   Examples include $S(P, y)=-\log\mathrm{det} (\Sigma_P)- (y- \mu_P )^T   \Sigma_p^{-1}(y-\mu_P )$, where $\mu_P $ and $\Sigma_P$ are the mean vector and covariance matrix of distribution $P$.
\end{enumerate}

Now return to the problem of model combination after specifying the score rule $S$ and corresponding divergence $d$.  The observed data are $y=(y_1, \dots , y_n)$. For simplicity, we remove all covariates $x$ in the notation.  Suppose we have  a set of probabilistic models $\mathcal{M}= ( M_1, \dots, M_K)$; then the goal in stacking is to find an optimal super-model in the convex linear combination  with the form  $\mathcal{C}= (\sum_{k=1}^{K}w_k p(\cdot | M_k) |   \sum_k w_k=1, w_k \geq 0  )$, such that its divergence to the true data generating model, denoted by $p_t(\cdot|y)$, is minimized: 
$$\min_{w}  d \Bigl( \sum_{k=1}^K w_k p( \cdot | y, M_k ), p_t(\cdot | y) \Bigr) . $$
Or  equivalently maximize the scoring rule of the predictive distribution,
\begin{equation} \label{stacking_population}
\max_{w}  S\Bigl(   \sum_{k=1}^K w_k p(\cdot | y, M_k), p_t(\cdot | y)   \Bigr),  
\end{equation}
where $ p( \tilde y | y, M_k)$ is the predictive density of $\tilde y$ in model $k$:
$$
p( \tilde y | y, M_k) = \int  p( \tilde y | \theta_k, y, M_k)p(\theta_k | y, M_k)d \theta_k  .
$$
We label the leave-$i$-out predictive density in model $k$ as,
$$\hat p_{k, -i}(y_i)= \int p(y_i | \theta_k, M_k) p(\theta | y_{-i}, M_k)  d\theta_k . $$
where $y_{-i}=(y_1,\dots, y_{i-1}, y_{i+1}, \dots,  y_n)$.

Then we define the stacking weights as the solution to the following optimization problem:
\begin{equation} \label{stacking}
 \max_w \frac{1}{n}\sum_{i=1}^n S\Bigl( \sum_{k=1}^K  w_k \hat p_{k,-i}, y_i\Bigr) , \quad  s.t. \quad w_k \geq 0,   \quad \sum_{k=1}^K w_k=1.
\end{equation}
Eventually, the combined estimation of  the predictive density is
\begin{equation} \label{predictive}
 \hat p(\tilde y |y)= \sum_{k=1}^K \hat w_k p(\tilde y|y, M_k).
 \end{equation}
When using logarithmic score (corresponding to Kullback-Leibler divergence), we call this \emph{stacking of predictive distributions}:
\begin{equation*}
\max_w \frac{1}{n} \sum_{i=1}^n \log \sum_{k=1}^K w_k p(y_i | y_{-i}, M_k) , \quad s.t.   \quad w_k \geq 0,   \quad \sum_{k=1}^K w_k=1.
\end{equation*}
The choice of scoring rule can depend on the underlying application.  Stacking of means  (\ref{stackingMean}) corresponds to the energy score with $\beta = 2$. The reasons why we prefer stacking of predictive distributions (corresponding to  the logarithmic score) to stacking of means are: (i) the energy score with $\beta=2 $ is not a strictly proper scoring rule and can give rise  to identification problems, and (ii) every proper local scoring rule is equivalent to the logarithmic score \citep{score}.

\subsection{ Asymptotic behavior of stacking}

The stacking estimate  (\ref{stacking_population}) finds the optimal predictive distribution within the convex  set  $\mathcal{C}=\big\{  \sum_{k=1}^{K}w_k p(\cdot | M_k )   \mid    \sum_{k=1}^K w_k=1, w_k \geq 0  \big\} $, that is the closest  to the data generating process with respect to the chosen scoring rule. 
This is different from Bayesian model averaging, which asymptotically with probability 1 will select a single model:  the one that is closest in KL divergence to the true data generating process.

Solving for the stacking weights in (\ref{stacking})  is an M-estimation problem.   Under some mild conditions \citep{le2016bayes, clyde2013bayesian, key1999bayesian},  for either the logarithmic scoring rule or energy score (negative squared error)  and a given set of weights $w_1 \dots w_k$, as sample size $n \to \infty$, the following asymptotic limit holds:
$$
\frac{1}{n}\sum_{i=1}^n S\Bigl( \sum_{k=1}^K  w_k \hat p_{k,  -i}  , y_i \Bigr) -  \mathrm{E}_{\tilde y| y} S\Bigl(  \sum_{k=1}^K w_k p(\tilde y| y , M_k) , \tilde y\Bigr)          \xlongrightarrow{\text{$L_2$}}   0.
$$
Thus the leave-one-out-score is a consistent estimator of the posterior score. In this sense, the stacking weights is the optimal combination weights asymptotically.

In terms of \citet[][Section 3.3]{Vehtari+Ojanen:2012}, the proposed stacking with log score is $M_{*}$-optimal projection of the information in the actual belief model $M_{*}$ to $\hat{w}$, where explicit specification of $M_{*}$ is avoided by re-using data as a proxy for the predictive distribution of the actual belief model and $w_k$ are the free parameters.

\subsection{Pareto smoothed importance sampling}
One challenge in calculating the stacking weights proposed in (\ref{stacking}) is that we need the
leave-one-out (LOO) predictive density, 
$$p(y_i | y_{-i}, M_k)=\int p(y_i | \theta_k, M_k) p(\theta_k | y_{-i}, M_k) d\theta_k. $$ 
Exact LOO requires refitting each model $n$ times.  To avoid this onerous computation, we use the following approximate method.  For the $k$-th model, we fit to all the data, obtaining  simulation draws $\theta_k^s (s=1,\dots S)$ from the full posterior $p(\theta_k|y, M_k)$  and  calculate 
\begin{equation} \label{ratio}
r_{i,k}^s =\frac{1} {p(y_i | \theta^s, M_k) } \propto \frac{p(\theta^s | y_{-i}, M_k)}{ p(\theta^s | y,  M_k) }. 
 \end{equation}

The  ratio $ r_{i,k}^s$ has a density function in its denominator and can be unstable, due to a potentially long right tail. This problem can be resolved using Pareto smoothed importance sampling (PSIS).  For each fixed model $k$ and data $i$, we fit the generalized Pareto distribution to the 20\% largest importance ratios $ r_{i,k}^s$,  and calculate the expected values of the order statistics of the fitted generalized Pareto distribution. We further truncate those values to get the smoothed importance weight $w_{i,k}^s$, which is used to replace $ r_{i,k}^s$. For details of PSIS, see \cite{practicalPSIS}. In the end, the LOO importance sampling is performed using
  $$p(y_i | y_{-i}, M_k )  \approx   \frac{1}{ \frac{1}{S}  \sum_{s=1}^{S} w_{i,k}^s  } .$$

When stacking using the logarithmic score, we are combining each model's   log predictive density. The PSIS estimate of the LOO expected log pointwise predictive density  in the $k$-th model is,
 $$ \mathrm { \widehat {elpd}_{loo}}^ k   =  \sum_{i=1}^n   \log\left( \frac{ \sum _{s=1}^S w_{i,k}^s p(y_i | \theta^s, M_k)   }{  \sum _{s=1}^S w_{i,k}^s } \right).$$
 
The reliability of the PSIS approximation can be determined by the estimated shape parameter $\hat k$ in the generalized Pareto distribution. For the left-out data points where $\hat k>0.7$, \cite{practicalPSIS} suggest replacing the PSIS approximation of those problematic cases by the exact LOO or $k$-fold validation. 

One potential drawback of LOO is the large variance when the sample size is small. We see in the simulation that when the ratio of relative sample size to effective number of parameters is small, the weighting can be unstable.  How to adjust this small sample behavior is left for the future research.

\subsection{Pseudo-BMA}\label{Pseudo-BMA}
In our paper, we also consider an AIC type weighting using leave-one-out cross-validation estimated expected log predictive densities. As mentioned in Section \ref{reference-bma}, these weights estimate the same quantities as \citet{dbd} that use the divergence from the reference model based inference. 

To maintain comparability with the given dataset and to get easier interpretation of the differences in scale of effective number of parameters, we define the expected log  pointwise predictive density (elpd) for a new dataset $\tilde y$ as a measure of predictive accuracy of a given model for the $n$ data  points  taken one at a time \citep{understanding}. In  model $M_k$, 
$$\mathrm{elpd}^k = \sum_{i=1}^{n} \int p_t(\tilde y_i) \log p(\tilde y_i |y, M_k) d \tilde y_i , $$
where $ p_t(\tilde y_i)$ denotes the true distribution of future data $\tilde y_i$. 

Given observed data y, we estimate elpd using LOO:
 $$ \widehat \mathrm{elpd}^k_{\mathrm{loo}} = \sum_{i=1}^{n}  \log p(y_i |y _{-i}, M_k) .  $$
Then we  define the Pseudo-BMA weighting for model $k$:
\begin{equation}\label{P-BMA}
w_k=  \frac{ \exp( \mathrm{\widehat {elpd}}^k_{\mathrm{loo}} ) }  { \sum_{k=1}^K \exp(  \mathrm{\widehat {elpd}}^{k}_{\mathrm{loo}} ) }.
\end{equation}
However, this estimation doesn't take into account the uncertainty resulted from having a finite number of proxy samples from the future data distribution.
Taking into account the uncertainty would regularize the weights
making them go further away from 0 and 1.

The computed estimate elpd is defined as the sum of $n$ independent components so it  is trivial to compute their standard errors by computing the standard deviation of the $n$ pointwise values \citep{vehtari2002bayesian}.  Define  
$$\mathrm{\widehat {elpd}}_{\mathrm{loo},i}^k=\log p(y_i | y_{-i }, M_k),$$ 
and then we can calculate
$$\mathrm{se} (   \mathrm{\widehat {elpd}}_{\mathrm{loo},i}^k )  =  \sqrt{  \sum_{i=1}^n    (\mathrm{\widehat {elpd}}_{\mathrm{loo},i}^k- \mathrm{\widehat {elpd}}_{\mathrm{loo}}^k /n  )^2  }.$$
Simple modification of weights is to use the lognormal approximation:
$$w_k=  \frac{ \exp(  \mathrm{\widehat {elpd}}^k_{\mathrm{loo}}   -\frac{1}{2} \mathrm{se (\widehat {elpd}}_{\mathrm{loo}}^k )  )}  { \sum_{k=1}^K    \exp(\mathrm{\widehat {elpd}}^k_{\mathrm{loo}}     -\frac{1}{2} \mathrm{se (\widehat {elpd}}_{\mathrm{loo}}^k )       ) }.$$
Finally, Bayesian bootstrap (BB) can  be used to compute uncertainties related to LOO estimation \citep{vehtari2002bayesian}.  Bayesian bootstrap \citep{rubin1981bayesian} makes simple non-parametric approximation to the distribution of random variable. Having samples of $z_1,\dots , z_n$ from a random variable $Z$, it is assumed that posterior probabilities for all observed $z_i$ have the distribution $\mbox{Dirichlet}(1,\dots,1)$ and values of $Z$ that are not observed have zero posterior probability. Sampling from the uniform Dirichlet distribution gives BB samples from the distribution of  $Z$ and thus samples of any parameter of this distribution can be obtained. 
In other words, each BB replication generates a set of posterior probability $\alpha_{1:n}$ for all observed $z_{1:n}$,
$$  \alpha_{1:n}\sim \mbox{Dirichlet}(\overbrace{1,\dots, 1}^{n}),  \quad P(Z=z_i |\alpha)=\alpha_i. $$
This leads to one BB replication of any statistic $\phi(Z)$ that is of interest:
$$\hat \phi(Z | \alpha) = \sum_{i=1}^n \alpha_i \phi(z_i).  $$
The distribution over all replicated $\hat \phi(Z |\alpha)$ (i.e., generated by repeated sampling of $\alpha$) produces an estimation for $\phi(Z)$.

As the distribution of $\mathrm{\widehat {elpd}}^k_{\mathrm{loo},i  }$ is often highly skewed, BB is likely to work better than the Gaussian approximation. In our model weighting, we can define
$$z_i^k = \mathrm{\widehat {elpd}}^k_{\mathrm{loo},i}, \quad i=1,\dots n.$$
We sample vectors $(\alpha_{1,b},\dots ,\alpha_{n,b})_{b=1,\dots ,B}$ from the Dirichlet $(\overbrace{1,\dots, 1}^{n})$ distribution, and compute the weighted means,
$$ \bar z_b^k = \sum_{i=1}^n \alpha_{i,b}z_i^k $$
Then a Bayesian bootstrap sample of $w_k$ with size $B$ is,
$$w_{k,b}=  \frac{ \exp(n  \bar{z}^k_b  ) }{\sum_{k=1}^K\exp(n  \bar{z}^{k}_b  )},\quad {b=1,\dots ,B}$$
and the final adjusted  weight of model $k$ is,
\begin{equation}
 w_{k}= \frac{1}{B}\sum_{b=1}^B w_{k,b},
 \end{equation}
which we call Pseudo-BMA+ weights.

\section{Simulation examples}

In this section, we first illustrate the advantage of  stacking of predictive distributions with a Gaussian mixture model. Then we compare stacking, BMA, Pseudo-BMA, Pseudo-BMA+ and other averaging methods through a series of linear regression simulation, where stacking gives the best performance in most cases. Finally we apply stacking to two real datasets, averaging multiple models so as to better explain US Senate voting or well-switching pattern in Bangladesh. 

\subsection{Gaussian mixture model}

This simple example helps us understand how BMA and stacking behave differently. It also illustrates the importance of the choice of the scoring rules in combining distributions.  Suppose the observed data $y=(y_i, i=1,\dots ,n)$  come iid from a normal distribution N$(3.4, 1)$, not known to the data analyst, and there are 8 candidate models, $\mbox{N}(1, 1)$, $\mbox{N} (2, 1)$,\dots , $\mbox{N} (8, 1)$. This is an $\mathcal{M}$-open problem in that none of the candidates is the true model, and we have set the parameters so that the models are somewhat separate but not completely distinct in their predictive distributions.

For BMA with a uniform prior $\mbox{Pr}(M_k)= \frac{1}{8}, k=1,\dots ,8$, we can write the posterior distribution explicitly:
$$\hat w_k^{\mathrm{BMA}}=P(M_k | y) =   \frac{\exp(  -  \frac{1}{2}\sum_{i=1}^n (y_i- \mu_k)^2 )}{ \sum_{k'} \exp(-  \frac{1}{2} \sum_{i=1}^n (y_i- \mu_{k'})^2 ) },
$$
from which we see that $\hat w_3^{\mathrm{BMA}} \xlongrightarrow{\text{$P$}}  1$ and $\hat w_k^{\mathrm{BMA}} \xlongrightarrow{\text{$P$}}  0$ for   $k\neq 3$ as sample size $n \to\infty $.
Furthermore, for any given $n$ ,
\begin{eqnarray*}
\mathrm {E}_{y\sim \mathrm{N}(\mu, 1)}[ \hat w_k^{\mathrm{BMA}}]  &\propto&  E_y \left(\exp(-  \frac{1}{2} \sum_{i=1}^n (y_i- \mu_k)^2 ) \right) \\
&\propto& \left(\int_{-\infty}^{\infty}  \!\!\exp \left(-  \frac{1}{2}\left(  (y- \mu_k)^2 +(y-\mu)^2 )\right) \right) dy\right)^n\\
&\propto& \exp\left(- \frac{  n(\mu_k-\mu)^2}{4}\right),
\end{eqnarray*}
where $\mu=3.4$ and $\mu_k=k$ in this setting.

  \begin{figure}
 \includegraphics[width=\textwidth ] {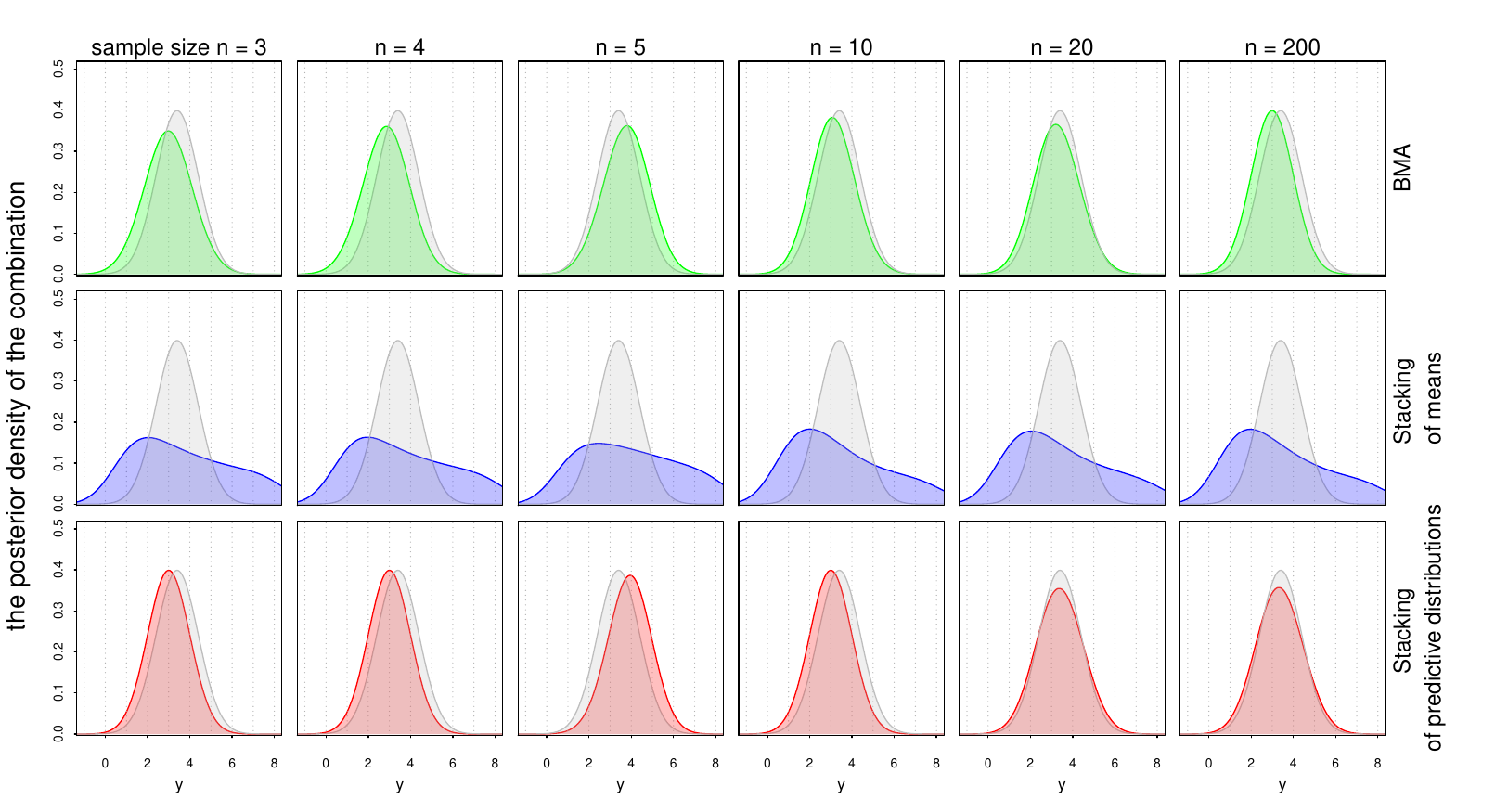}
\vspace{-.3in}
\caption{ \em For the Gaussian mixture example, the predictive distribution $p( \tilde y |y)$ of BMA (green curve), stacking of means (blue) and stacking  of predictive distributions (red). In each graph, the gray distribution represents the true model $\mbox{N}(3.4,1)$.  Stacking of means matches the first moment but can ignore the distribution. For this $\mathcal{M}$-open problem, stacking of predictive distributions outperforms BMA as sample size increases. } \label{dis34}
 \end{figure}

This example is simple in that there is no parameter to estimate within each of the models: $p( \tilde y | y , M_k)=p( \tilde y |   M_k)$.  Hence, in this case the weights from Pseudo-BMA and Pseudo-BMA+ are the same as the BMA weights, $\exp(\mathrm{\widehat{elpd}}_{\mathrm{loo}}^k)/\sum_{k'} \exp(\mathrm{\widehat{elpd}}_{\mathrm{{loo}}}^{k'}) $. 

For  stacking of means,  we need to solve
$$\hat w =\arg \min_{w} \sum_{i=1}^n (y_i-  \sum_{k=1}^8 w_k k  )^2  , \quad s.t.   \sum_{k=1}^8 w_k=1,  \quad w_k \geq 0.$$
This is nonidentifiable because the solution contains any vector $\hat w$ satisfying $$\sum_{k=1}^8 \hat w_k=1, \quad \hat w_k \geq 0, \quad \sum_{k=1}^{8} \hat w_k k = \frac{1}{n}\sum_{i=1}^n y_i.$$ For point prediction, the stacked prediction is always  $\sum_{k=1}^{8} \hat w_k k = \frac{1}{n}\sum_{i=1}^n y_i$, but it can lead to different predictive distributions $\sum_{i=1}^k \hat w_i \mbox{N}(k,1)$.  To get one reasonable result, we transform the least square optimization to the following normal model and assign a uniform prior to $w$: 
 $$  y_i\sim \mbox{N}\left(\sum_{k=1}^8 w_k k  , \sigma^2\right), \quad   p(w_1,\dots ,w_8, \sigma )=1 . $$
Then we could use the posterior means  of $w$ as model weights.
 
For stacking of predictive distributions,we need to solve
$$\max_w  \sum_{i=1}^n \log \left(  \sum_{k=1}^8 w_k \exp\left(-\frac{(y_k - k)^2}{2}  \right)\right),  \quad s.t.   \sum_{k=1}^8 w_k=1,  \quad w_k \geq 0 $$

In fact, this example is a density estimation problem.  \cite{smyth1998stacked}  first apply stacking to non-parametric density estimation, which they call \emph{stacked density estimation} and now can be viewed as a special case of our stacking method. 

  \begin{figure}
 \includegraphics[width=\textwidth ] {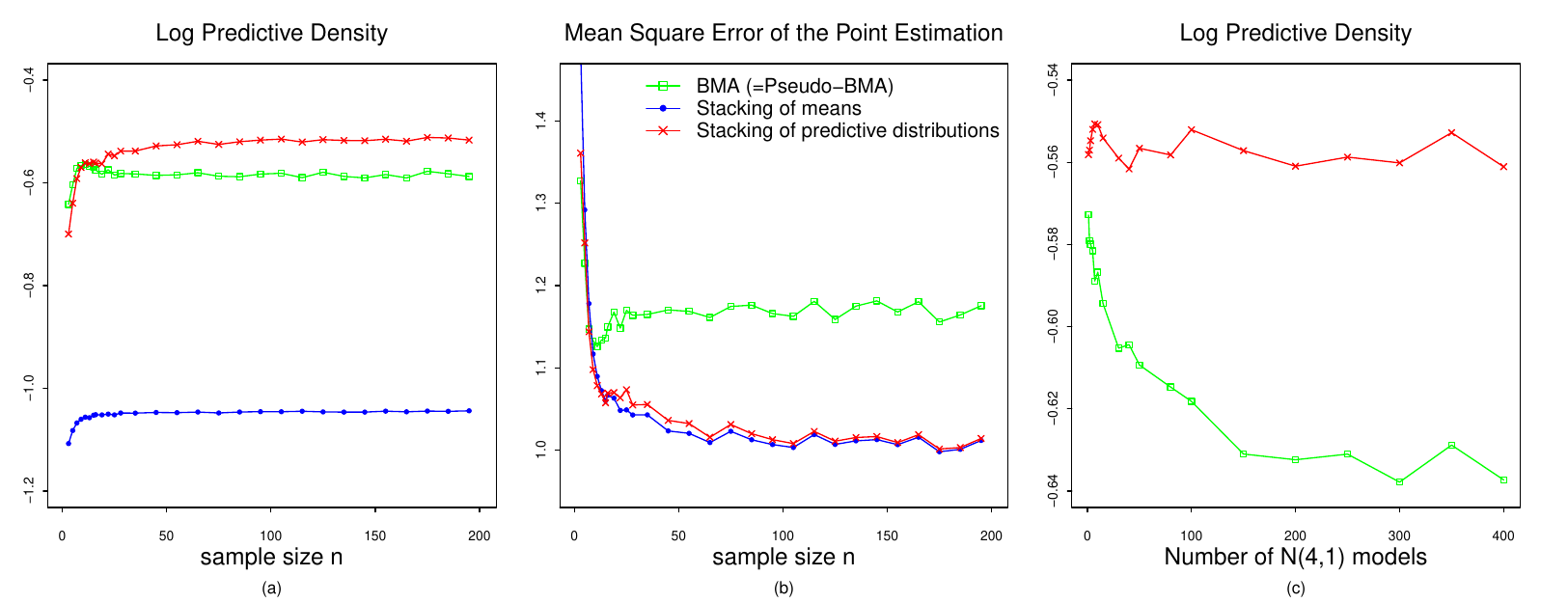}
 \vspace{-.3in}
 \caption{\em (a) The left panel  shows the expected log predictive density  of the  predictive distribution under  BMA , stacking of means and stacking  of predictive distributions. Stacking of predictive distributions performs best for moderate and large sample sizes.  (b) The middle panel shows  the  main squared error treating the posterior mean of $\hat y$ as a point estimation. Stacking  of predictive distributions gives almost the same optimal mean squared error as stacking of means, both of which perform better than  BMA.  (c) The right panel shows the expected log predictive density of stacking and BMA when adding some more {N}$(4,1)$ models to the model list, where sample size is fixed to be 15. All average log scores and errors are calculated through 500 repeated simulation and 200 test data generating from the true distribution.} \label{score34}
 \end{figure}

We  compare the posterior predictive distribution $ \hat p( \tilde y | y)  = \sum_k \hat w_k   p( \tilde y | y , M_k) $, for these three methods of model averaging.  Figure  \ref{dis34}  shows the predictive distributions in one simulation when the sample size $n$ varies from 3 to 200.   We first notice that stacking  of means (the middle row of graphs) gives an unappealing predictive distribution, even if its point estimate is reasonable.  The broad and oddly spaced distribution here arises from nonidentification  of $w$, and it demonstrates the general point that stacking of means does not even try to match the shape of the predictive distribution. The top and bottom row of graphs show how BMA picks up the single model that is closest in KL divergence, while stacking picks a combination; the benefits of stacking becomes clear for large $n$. 

In this trivial non-parametric case, stacking  of predictive distributions is  almost the same as fitting a mixture model, except for the absence of the prior.  The true model N$(3.4, 1)$ is actually a convolution of single models rather than a mixture, hence no approach can recover the true one from the model list.   From Figure \ref{score34} we can compare the  mean squared error and the mean logarithmic score of this three combination methods. The average log scores and errors are calculated through 500 repeated simulation and 200 test data generating from the true distribution.  The left panel  shows logarithmic score  (or equivalent, expected log predictive density) of the  predictive distribution. Stacking  of predictive distributions always gives a larger score except for extremely small $n$.  
In the middle panel, it shows the  main squared error by considering the posterior mean of predictive distribution to be a point estimation, even if it is not our focus.  In this case, it is not surprising to see that stacking  of predictive distributions gives almost the same optimal mean squared error as the stacking of means, both of which are better than the BMA.  Two distributions close in KL divergence are close in each moment, while the reverse doesn't necessarily hold. This illustrates the necessary of  matching the \emph{distribution}, rather than matching the \emph{moments} for the predictive distribution.

Finally it is worth pointing out that stacking depends only on the space expanded by all candidate models, while BMA or Pseudo-BMA weighting may by misled by such model expansion. If we add another  N$(4, 1)$ as the $9$th model  in the model list above, stacking will not change at all in theory, even though it becomes non-strictly-convex and has infinite same-height mode. For BMA, it is equivalent to putting double prior mass on the original $4$th model, which doubles the final weights for it.  The right panel of Figure  \ref{score34} shows such phenomenon: we fix sample size $n$ to be 15 and add more and more N$(4,1)$ models. As a result, BMA (or Pseudo-BMA weighting) puts more weights on N$(4,1)$ and behaves worse, while stacking changes nowhere except for numerical fluctuation. It illustrates another benefit of stacking compared to BMA or Pseudo-BMA weights. We would not expect a combination method that would  performs even worse as model candidate list expands, which may become a disaster when many similar weak models exist. We are not saying BMA can never work in this case. In fact some other methods are proposed to make BMA overcome such drawbacks.  For example, \cite{george2010dilution} establishes dilution priors to compensate for model space redundancy for linear models, putting less weights on those models that are close to each other.  \cite{fokoue2011bias} introduce prequential model list selection to obtain an optimal model space. But we propose stacking as a more straightforward solution.

\subsection{Linear subset regressions}\label{reg}
The previous section demonstrates a simple example of combining several different nonparametric models.  Now we turn to the  parametric case. This example comes from \cite{breiman1996} who compares stacking to model selection. Here we work in a Bayesian framework.

 Suppose the true model is 
 $$Y=\beta_1X_{1}+\dots +\beta_JX_{J}+\epsilon.$$
 In the model $\epsilon$  independently comes  from N$(0, 1)$.  All the covariates  $X_{j}$ are independently  from N$(5, 1)$. The number of predictors $J$ is 15. 
The coefficient $\beta $ is generated by
$$ \beta_j=\gamma\left(     (1_{ | j-4 | <h}  (h- |j-4|)^2  +(  1_{|j-8|<h }   )  (  h- |j-8|  )^2+ (   1_{|j-12|<h}    ) (  h-|j-12|  )^2 \right), $$
where $\gamma$ is determined by fixing the signal-to-noise ratio such that   
$$\frac{\mathrm{Var}(\sum_j \beta_j X_j ) }{1+\mathrm{Var}(\sum_j \beta_j X_j )}=\frac{4}{5}. $$
The value $h$ determines the number of nonzero coefficients in the true model. For $h=1$, there are 3 ``strong" coefficients. For $h=5$, there are 15 ``weak" coefficients. In the following simulation, we fix $h=5$.

We consider the following two cases:

  \begin{figure}
   \vspace{-.1in}
 \centerline{\includegraphics[width=.9\textwidth ] {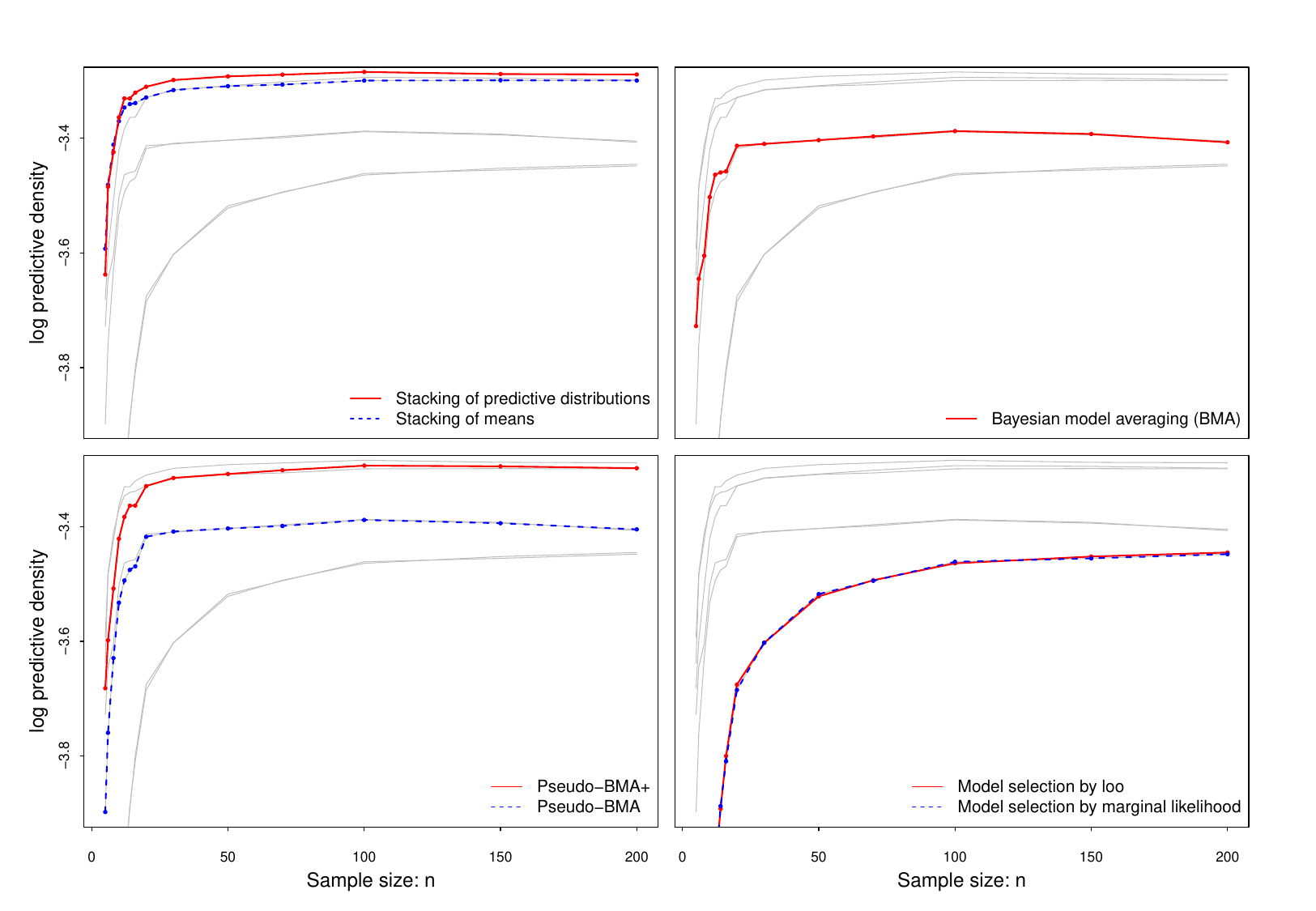}}
\caption{\em  Mean log predictive density  of 7 combination methods    in  the linear regression variable selection example: the $k$-th model is a univariate linear regression with the $k$-th variable.    The mean log score is evaluated by 100 repeated experiments and  200 test data. } \label{univariate}
   \vspace{-.1in}
\end{figure}

\begin{enumerate}
\item $\mathcal{M}$-open

Each subset contains only one single variable. Hence, the $k$-th model is a univariate linear regression with the $k$-th variable $X_k$. We have $K=J=15 $ different  models in total. One advantage of stacking and Pseudo-BMA weighting is that they are  not sensitive to prior, hence even a flat prior will work, while BMA can be sensitive to the  prior. For each single model $M_k: Y \sim  \mbox{N}(\beta_k X_k ,  \sigma^2) $, we set prior $\beta_k \sim N(0,10)$, $\sigma \sim \mathrm{Gamma} (0.1,0.1) $. 

\item  $\mathcal{M}$-closed

Let model $k$  be the linear regression with subset $(X_1,\dots, X_k)$. Then there are still $K=15$ different models.  Similarly, in model   $M_k: Y \sim \mbox{N}( \sum_{j=1}^k \beta_j X_j , \sigma^2) $, we set prior $\beta_j \sim N(0,10), j=1,\dots, k$, $\sigma \sim \mathrm{Gamma} (0.1,0.1) $.
\end{enumerate}

In both cases, we have seven methods for combine predictive densities: (1) stacking of predictive distributions, (2) stacking of means, (3) Pseudo-BMA, (4) Pseudo-BMA+, (5) best model selection by mean LOO value, (6) best model selection by marginal likelihood, and  (7) BMA.   A linear combination $p( \tilde y | y)= \sum_{k=1}^K \hat w_k p( \tilde y | M_k) $ is what we have in the end as  estimation for the posterior density of the new data $\tilde y$. 
 We generate a test dataset $(\tilde x_i ,\tilde y_i)$, $i=1,\dots,200$ from the underlying true distribution  to calculate the out of sample score for the combined distribution under each method $k$:  $$\frac{1}{200}\sum_{i=1}^{200}  \log \sum_{k=1}^K \hat w_k p( \tilde y_i | M_k) .$$  We loop the test simulation 100 times to get  the expected predictive accuracy for each method.

 \begin{figure}
    \vspace{-.1in}
\centerline{ \includegraphics[width=.9\textwidth ] {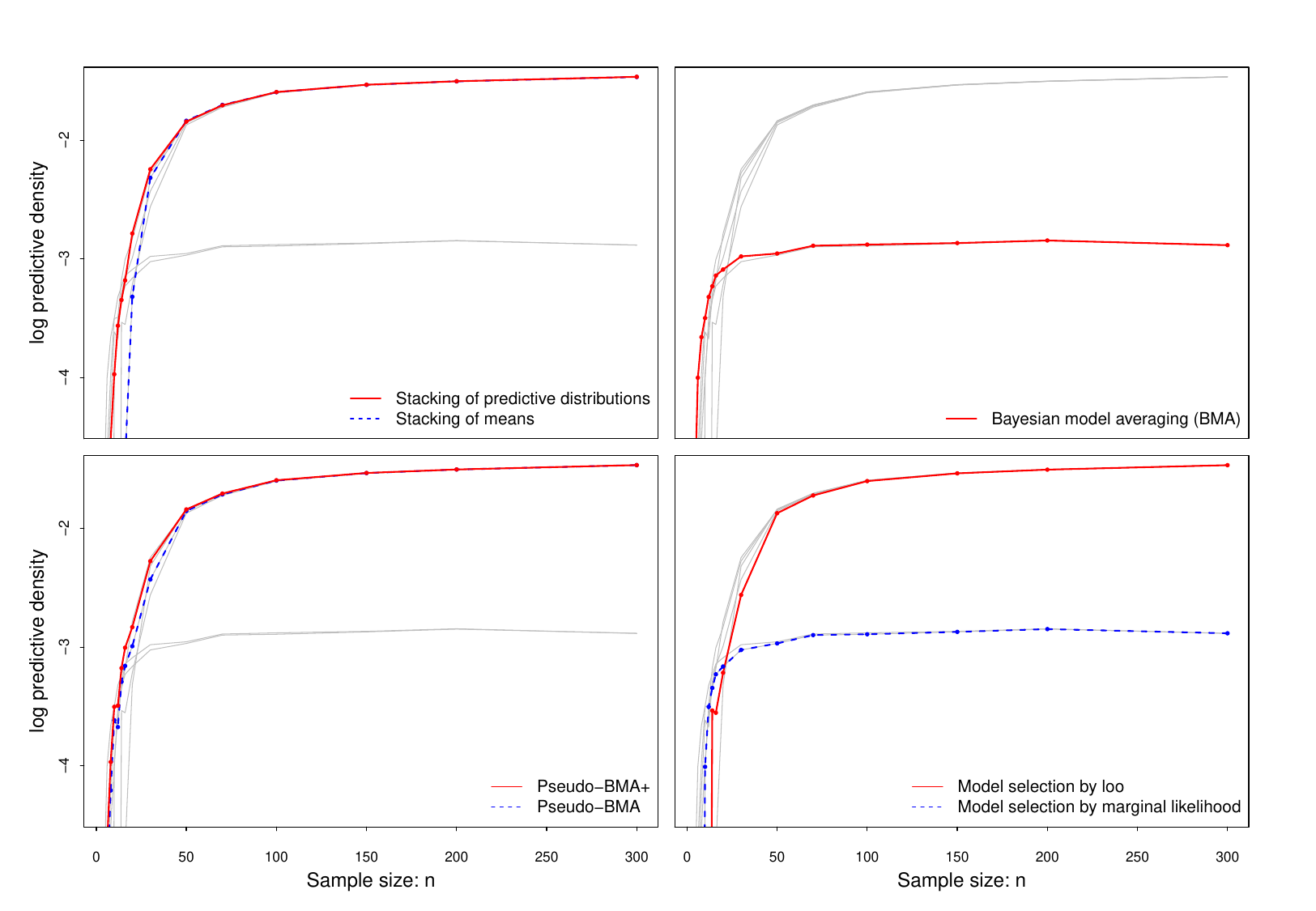}}
\caption{\em Mean log predictive density  of 7 combination methods in the   linear regression variable selection example:  the $k$-th model is a linear regression with the first $k$ variables.  We evaluate the mean log score using 100 repeated experiments  and 200 test data. } \label{regression}
    \vspace{-.1in}
 \end{figure}

Figure \ref{univariate} shows the expected out-of-sample log predictive density for the seven methods, for a set of experiments with sample size $n$ ranging from 5 to 200.   Stacking seems to outperform all other methods even for small $n$. Stacking of predictive distributions is asymptotically better than any other combination method. Pseudo-BMA+ weighting dominates naive Pseudo-BMA weighting.  Finally, BMA performs similarly to Pseudo-BMA weighting, always better than any kind of model selection, but that advantage vanishes in the limit since BMA picks up one model.  In this $\mathcal{M}$-open setting,  we know model selection can never be optimal.

The results change when we move to the second case, in which the $k$-th model contains variables $X_1,\dots, X_k$ so that we are comparing models of differing dimensionality.  The problem is $\mathcal{M}$-closed because  the largest subset contains all the variables, and we have simulated data from this model.  Figure \ref{regression} shows the mean log predictive density of the seven combination methods in this case. For a large sample size $n$, almost all methods recover the true answer (putting weight 1 on the full model), except BMA and model selection based on marginal likelihood. The poor performance of BMA comes from the parameter priors:  recall that the optimality of BMA arises when averaging over the priors and not necessarily conditional on any particular chosen set of parameter values.  There is no  general no way to obtain a ``correct'' prior that accounts for the complexity for BMA in an arbitrary model space. Model selection by LOO can recover the true model, while selection by marginal likelihood cannot due to the same prior problems.  Once again, we see that BMA eventually become the same as model selection by marginal likelihood, which is much worse than any other methods asymptotically.

In this example, stacking is unstable for extremely small $n$. In fact, our computations for stacking of predictive distributions and Pseudo-BMA depend on the the PSIS approximation $\log p(y_i | y_{-i})$. If this approximation is crude, then the second step optimization cannot be accurate.   It is known that the parameter $\hat k$ in  the generalized Pareto distribution can be used to diagnose the accuracy of PSIS approximation. In our method, we replace PSIS approximation  by running exact LOO for any data points with estimated  $\hat k > 0.7$ \citep[see ][]{practicalPSIS}. 

 \begin{figure}
 \includegraphics[width=\textwidth ] {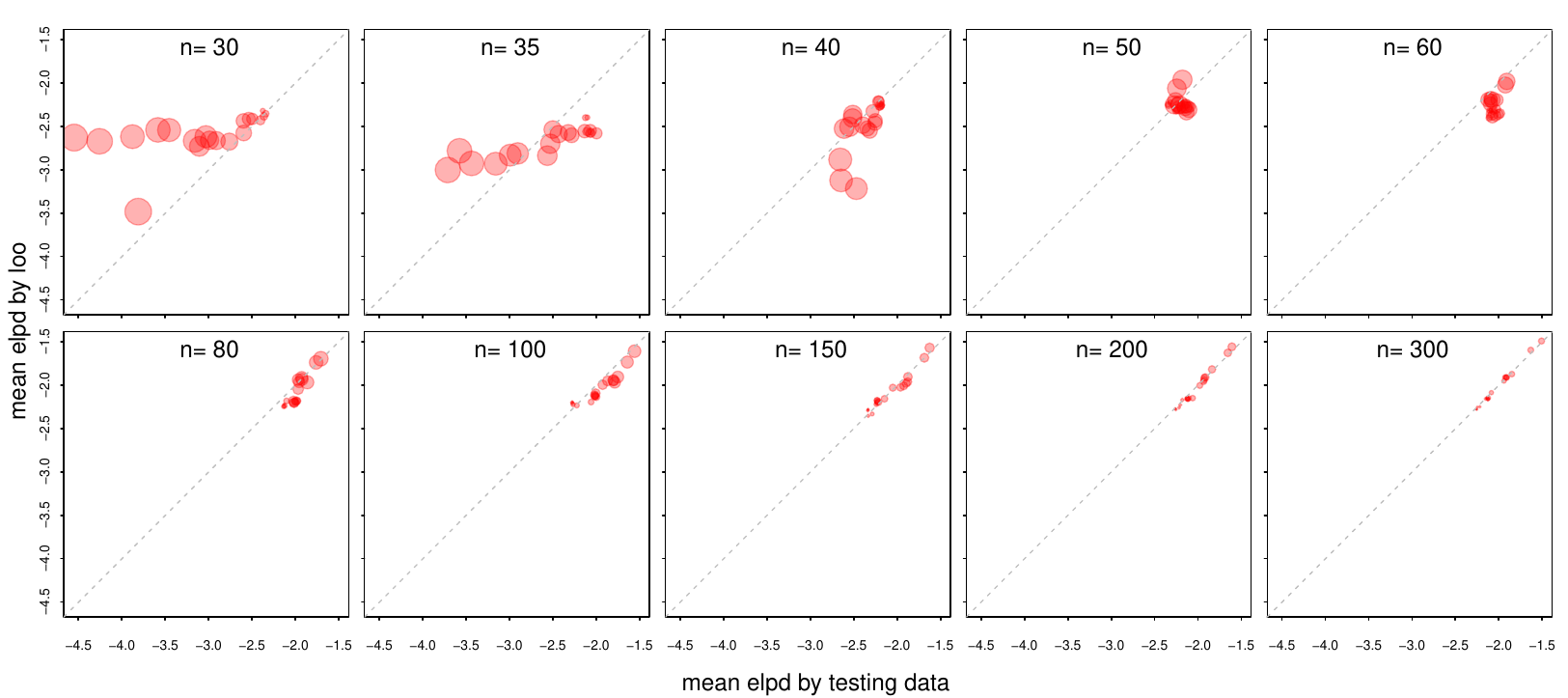}
\caption{\em Comparison  of the mean elpd  estimated by LOO and the mean elpd calculated from test data, for each model and each sample size in the simulation described above. The area of each dot represents the relative complexity of the  model as measured by  effective number of parameter divided by sample size.    } \label{PSIS}
 \end{figure}

Figure \ref{PSIS} shows the comparison of the mean elpd estimated by LOO, $$\frac{1}{n} \mathrm{elpd}_{\mathrm{loo}}^k = \frac{1}{n} \sum_{i=1}^n\log    p_k (y_i | y_{-i}),$$  and the mean elpd calculated using 200  independent test data,
 $$ \frac{1}{n} \mathrm{elpd}_{\mathrm{test}}^k= \frac{1}{200} \sum_{i=1}^{200} \log    p_k (\tilde y_i  | y)  $$ for each model $k$ and each sample size in the simulation described above. The area of each dot in Figure \ref{PSIS} represents the relative complexity of the model as measured by effective number of parameters in the model divided by sample size.   We evaluate the effective number of parameters  using LOO \citep{practicalPSIS}.    The sample size $n$ varies from 30 to 200 and variable size is fixed to be 20.   Clearly, the relationship is far from the  line $y=x$ for extremely small sample size, and the relative bias ratio ($\mathrm{elpd_{loo}} / \mathrm{elpd_{test}} $) depends on the complexity of the model.   Empirically, we have found the approximation to be poor when the sample size is less than 5 times the number of parameters.

As a result, in the small sample case, LOO can lead to relatively  large variance, which makes the stacking of predictive distributions  and Pseudo-BMA/ Pseudo-BMA+  unstable, with performance improving quickly as $n$ grows.

\subsection {Comparison with  mixture models}

Stacking is inherently a two-step procedure.  In contrast, when fitting a mixture model, one estimates the model weights and the status within parameters in the same step.  In a mixture model, given a model list $\mathcal{M}=\{M_1, \dots, M_k\}$, each component in the mixture occurs with probability $w_k$. Marginalizing out the discrete assignments yields the  joint likelihood
$$ p(y |w_{1:K}, \theta_{1:K}  ) =\sum_{k=1}^K  w_k p(y | \theta_k , M_k)  . $$


The mixture model seems to be the most straightforward  continuous model expansion.  Nevertheless, there are several reasons why we may prefer stacking to the mixture model, though  the latter one is a full Bayesian approach.     First, the computation  cost of  mixture models can be relatively large.   If the true model is a mixture model and the estimation of each model depends a lot on others, then it is worth paying the extra computation cost. However,  it is not quite possible to  combine several components in real application.  It is more likely that the researcher is running combination among several mixture models with different pre-specified number of components.  The model space can be always extended so it is infeasible to make such kind of full Bayesian inference. 
 
Second, if every single model is close to one another and sample size is small, the mixture model can face non-identification or instability problem, unless a strong prior is added.  Since the mixture model is relatively complex, this leads to a poor small sample behavior.

\begin{figure}
\vspace{-.2in}
 \includegraphics[width=\textwidth ] {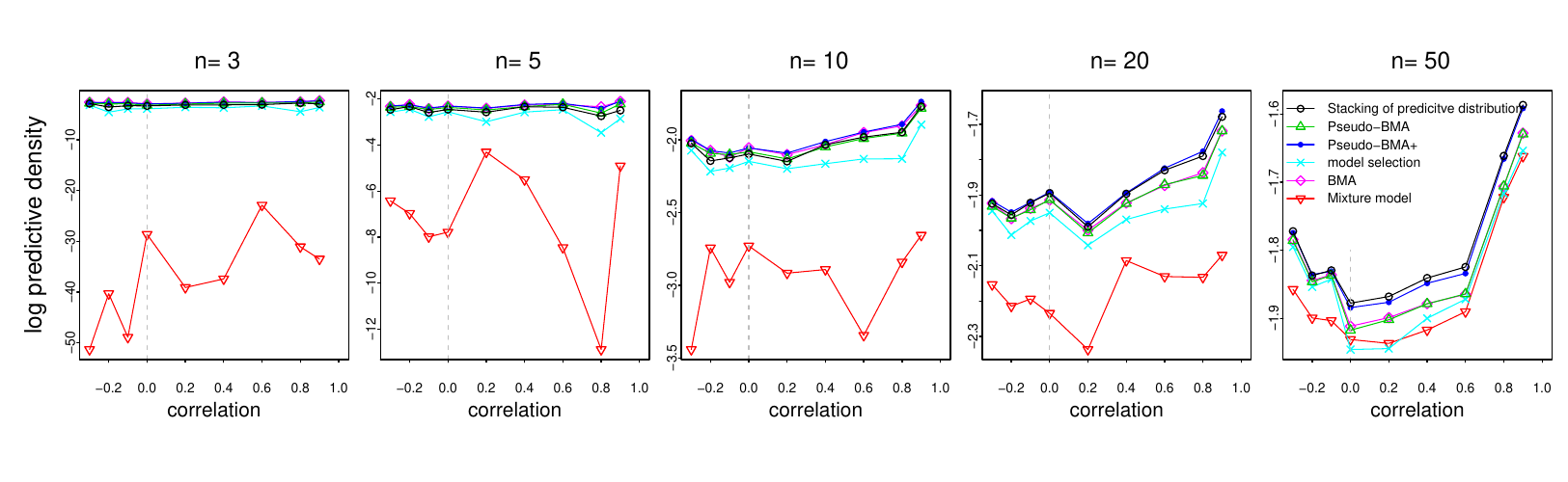}
\vspace{-.4in}
\caption{\em Log predictive density  of the combined posterior distribution  obtained by stacking of predictive distributions, BMA, Pseudo-BMA,  Pseudo-BMA+, model selection by marginal likelihood, or mixture models. In each case, we evaluate the  predictive density by taking mean of 100 testing data and 100 repeated simulations experiments.  The correlation  of variables range from $-0.3$ to 0.9, and sample sizes range from 4 to 50. Stacking on predictive distribution and Pseudo-BMA+ outperform mixture models in all cases. } \label{mixture}
 \end{figure}
 
Figure \ref{mixture} shows a comparison of mixture model and our model averaging methods in a numerical experiment, in which the true model is
 $$Y \sim \mbox{N}(  \beta_1 X_1 + \beta_2 X_2 + \beta_3 X_2 ,1), \quad \beta_k  \text{ is generated from  } \mbox{N}(0,1),$$
and there are 3 candidate models, each containing one covariate:
 $$M_k: Y \sim \mbox{N}(  \beta_k X_k ,\sigma_k^2),  \text{with a prior } \beta_k \sim \mbox{N}(0,1) ,  \quad k=1,2,3 .$$

 In the simulation, we generate the design matrix by $\mathrm{Var}(X_i)=1$ and $\mathrm{Cor}(X_i, X_j)=\rho$.  $\rho$ determines how correlated these models are and it ranges from $-0.3$ to 0.9. 
 
Figure \ref{mixture} shows that both the performance of mixture models and single model selection are worse than any other model averaging method we suggest, even though the mixture model takes much longer time to run (about 30 more times) than stacking or  Pseudo-BMA+.   When the sample size is small, the mixture model is too complex to fit. On the other hand, stacking  of predictive distributions and  Pseudo-BMA+ outperform all other methods with a moderate sample size.

 \cite{clarke2003} argues that the effect of (point estimation) stacking only depends on the space spanned by the model list, hence he suggests  putting those ``independent" models in the list.  Figure \ref{mixture} shows high correlations do not hurt stacking and  Pseudo-BMA+ in this example. 

\subsection{Variational inference with different initial values}
In Bayesian inference, the posterior density of parameters $\theta=\{\theta_1, \dots, \theta_m \}   $ given observed data $y=\{y_1, \dots, y_n \}$ can be difficult to compute. Variational inference   can be used to  give a fast approximation for  $p(\theta | y)$ \citep[for a recent review, see][]{blei2017variational}.  Among a family of distribution $\mathcal{Q}$,  we try to  find the member of that family such that the Kullback-Leibler divergence to the true distribution is minimized:
 \begin{equation} \label{VI}
 q^*(\theta) = \arg_{q \in \mathcal{Q}} \min \mathrm{KL}\bigl(q(\theta), p(\theta | y) \bigr) = \arg_{q \in \mathcal{Q}} \min  \bigl( \mathrm{E}_q \log q(\theta)-\mathrm{E}_q \log p (\theta, y) \bigr),
 \end{equation}
   
One widely used variational family is mean-field family where parameters are assumed to be mutually independent $\mathcal{Q}= \{ q(\theta): q(\theta_1, \dots,\theta_m )=\prod_{j=1}^{m} q_j (\theta_j)  \}$.  Some recent progress is made to run  variational inference algorithm in a black-box way. For example, \citet{kucukelbir2016automatic} implement  \emph {Automatic Variational Inference} in Stan. Assuming all parameters $\theta$ are continuous and model likelihood is differentiable, it transforms $\theta$ into real coordinate space ${\rm I\!R}^m$ through  $\zeta=T(\theta)$ and  uses normal approximation $p(\zeta | \mu, \sigma^2)= \prod_{j=1}^m \hbox{N}(\zeta_j | \mu_j, \sigma_j^2 )$. Plugging this into  \ref{VI} leads to an optimization problem over $(\mu , \sigma^2) $, which can be solved by stochastic gradient ascent.  Under some mild condition, it eventually converges to a local optimum $  q^*(\theta)$.   However, $q^*(\theta)$ may depend on initialization since such optimization problem is in general non-convex,  particularly when the true posterior density $p(\theta|y )$ is multi-modal.

Stacking of predictive distributions and  Pseudo-BMA+ weighting can be used to  average several sets of posterior draws coming from different approximation distribution.  To do this, we repeat the variational inference $K$ times. At time $k$, we start from a random initial point and use  stochastic gradient ascent  to solve the optimization problem  \ref{VI}, ending up with an approximation distribution $q^*_k(\theta)$. Then we draw $S$ samples $\{\theta_k^{(1)}, \dots, \theta_k^{(S)}\}$ from $q^*_k$  calculate the importance ratio  $r_{i,k}^s$  defined in \ref{ratio} as 
$$r_{i,k}^s = \frac{1}{p( y_i | \theta ^{(s)}_{k} )} $$
After this,  the remaining steps follow as before.   We obtain  stacking or pseudo-BMA+ weights $w_k$ and average all  approximation distribution as $\sum_{k=1}^K w_k q^*_k $.

\begin{figure}
 \includegraphics[width=.9\textwidth ] {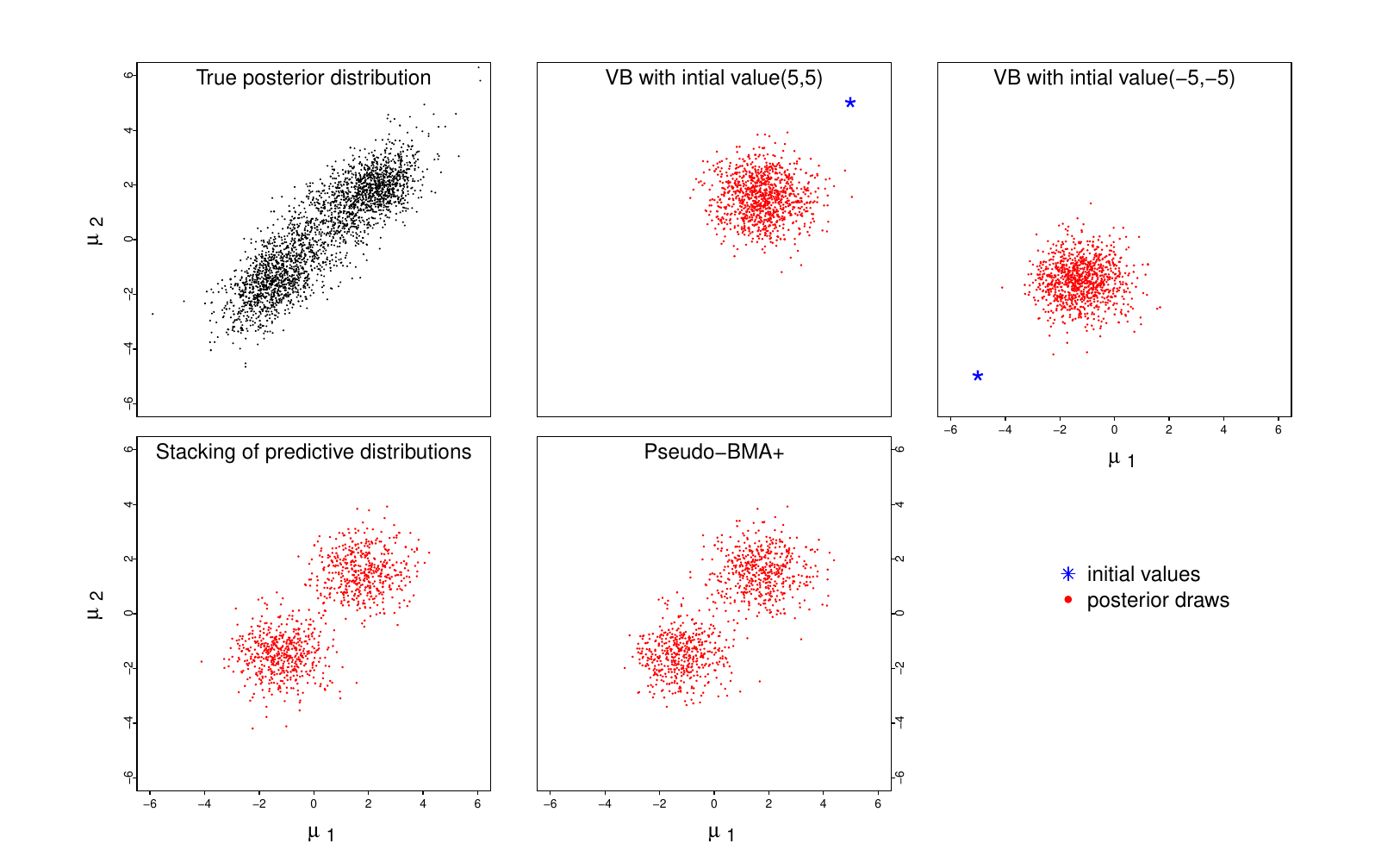}
\vspace{-.2in}
\caption{\em (1) A multi-modal posterior distribution of $(\mu_1, \mu_2)$.  (2--3) Posterior draws from variational inference with different initial values.  (4--5)  Averaged posterior distribution using stacking  of predictive distributions and Pseudo-BMA+ weighting. } \label{vb}
 \end{figure}
 
 Figure \ref{vb} gives a simple numerical example that the averaging strategy helps adjust the optimization uncertainty of initial values. Suppose the data is two-dimensional $y=(y^{(1)}, y^{(2)})$ and the parameter is $(\mu_1, \mu_2) \in {\rm I\!R}^2$. The likelihood $p(y | \mu_1, \mu_2)$ is given by 
$$y^{(1)} \sim \mathrm{Cauchy} (\mu_1,1), \quad y^{(2)} \sim \mathrm{Cauchy} (\mu_2,1). $$ 
A prior is assigned  on $\mu_1-\mu_2$:
$$\mu_1- \mu_2\sim \mbox{N}(0,1). $$
We generate  two observations $(y^{(1)}_{1}=3, y^{(2)}_{1}=2 )$ and $(y^{(1)}_{2}= -2, y^{(2)}_{2}=-2)$. The first panel shows  the true posterior distribution of $\mathbf{\mu}=(\mu_1, \mu_2)$, which is bimodal.  We run mean-field normal variational inference in Stan, with two initial values set to be $(\mu_1, \mu_2)=(5,5)$ and $(-5, -5)$ separately. This produces two distinct approximation distribution as shown in panel 2 and 3.  We then draw 1000 samples each from these two approximation distribution and use stacking or Pseudo-BMA+ to combine them. The lower 2  panels show the averaged posterior distribution. Though neither can recover the true distribution, the averaged version is closer to it, measured by KL divergence to the true one.

\subsection{Proximity and directional models of voting}

\cite{JOPO155} use  US Senate voting  data from 1988 to 1992 to study voters' preference on  the candidates who propose policies that are similar to them. They introduce  two similar variables that indicate the distance between voters and  candidates. \emph{Proximity voting comparison} represents the $i$-th voter's comparison between the candidates' ideological positions: 
$$ U_i(D) -U_i(R)= (x_R - x_i)^2 - (x_D - x_i)^2, $$
where $x_i$ represents the $i$-th voter's preferred ideological position, and $x_D$ and $x_R$ represent the ideological positions of the Democratic and Republican candidates, respectively.
In contrast, the $i$-th voter's \emph {directional comparison} is defined by
$$ U_i(D)-U_i(R)=(x_D -X_N)(x_i -X_N)-(x_R -X_N)(x_i -X_N),$$
where $X_N$ is the neutral point of the ideology scale. 

Finally, all these comparison is aggregated in the party level, leading to two party-level variable \emph{Democratic proximity  advantage} and \emph{Democratic directional advantage}. The sample size is $n=94$.

For both of these two variables, there are two ways to measure candidates' ideological positions $x_D$ and $x_R$, which leads to two different datasets. In the  \emph{Mean candidate} dataset, they are calculated by taking the average of  all respondents' answers in the relevant state and year. In the  \emph{Voter-specific} dataset,  they are calculate by  using respondents' own placements of the two candidates.    In both datasets, there are 4 other party-level variables. 

\begin{figure}
\centering
\resizebox{\columnwidth}{!}{%
\begin{tabular}{l cc cc cc cc}
                                                                           & \multicolumn{2}{c}{\textit{\textbf{Full model}}}  &  \multicolumn{2}{c}{\textit{\textbf{BMA}}}         & \multicolumn{2}{c}{\textit{\textbf{Stacking of}}}   & \multicolumn{2}{c}{\textit{\textbf{Pseudo-BMA+ weighting}}}   \\
                                                                           & \multicolumn{2}{c}{}  &  \multicolumn{2}{c}{}         & \multicolumn{2}{c}{\textit{\textbf{predictive distributions}}}   & \multicolumn{2}{c}{}   \\
                                                                           & \textit{Mean} & \textit{Voter-} & \textit{Mean} & \textit{Voter-} & \textit{Mean}      & \textit{Voter-}     & \textit{Mean}      & \textit{Voter-}      \\ 
 & \textit{Candidate} & \textit{specific} & \textit{Candidate} & \textit{specific} & \textit{Candidate}      & \textit{specific}     & \textit{Candidate}      & \textit{specific}      \\ \hline
\begin{tabular}{l}Dem.\\ prox.\\ adv.\end{tabular}  & -3.05 (1.32)          & -2.01 (1.06)           & -0.22 (0.95)          & 0.75 (0.68)           & 0.00 (0.00)                & 0.00 (0.00)   & -0.02 (0.08)                & 0.04(0.24)                \\
\begin{tabular}{l}Dem.\\ direct. \\ adv.\end{tabular} & 7.95 (2.85)           & 4.18 (1.36)           & 3.58 (2.02)           & 2.36 (0.84)           & 2.56 (2.32)                 & 1.93 (1.16)     & 1.60 (4.91)                & 1.78 (1.22)             \\
\begin{tabular}{l}Dem.\\ incumb.\\ adv.\end{tabular}  & 1.06 (1.20)           & 1.14 (1.19)           & 1.61 (1.24)           & 1.30 (1.24)           & 0.48 (1.70)                 & 0.34 (0.89)      & 0.66 (1.13)                & 0.54 (1.03)             \\
\begin{tabular}{l}Dem.\\ quality   \\ adv.\end{tabular}     & 3.12 (1.24)           & 2.38 (1.22)           & 2.96 (1.25)           & 2.74 (1.22)           & 2.20 (1.71)                 & 2.30 (1.52)     & 2.05 (2.86)                & 1.89 (1.61)             \\
\begin{tabular}{l}Dem.\\ spend\\ adv.\end{tabular}    & 0.27 (0.04)           & 0.27 (0.04)           & 0.32 (0.04)          & 0.31 (0.04)           & 0.31 (0.07)                 & 0.31 (0.03)      & 0.31 (0.04)                & 0.30 (0.04)            \\
\begin{tabular}{l}Dem.\\ partisan\\ adv.\end{tabular}    & 0.06 (0.05)           & 0.06 (0.05)           & 0.08 (0.06)          & 0.07 (0.06)          & 0.01 (0.04)                  & 0.00 (0.00)       & 0.03 (0.05)                & 0.03 (0.05)           \\
Const.      & 53.3 (1.2)           & 52.0 (0.8)           & 51.4 (1.0)           & 51.6 (0.8)          & 51.9 (1.1)                & 51.6 (0.7)      & 51.5 (1.2)                & 51.4 (0.8)           
\end{tabular}
}
\caption{\em Regression coefficient and standard error in the voting example, from the full model (columns 1--2), the subset regression model averaging using  BMA (columns 3--4), stacking of predictive distributions (columns 5--6) and Pseudo-BMA+ (columns 7--8).   \emph{Democratic proximity  advantage} and \emph{Democratic directional advantage}  are two highly correlated variables.  \emph{Mean candidate} and \emph{Voter-specific}  are two  datasets that provide different measurements on candidates' ideological placement. }
\label{Senate}
\end{figure}

The two variables \emph{Democratic proximity  advantage} and \emph{Democratic directional advantage} are highly correlated.  \cite{montgomery2010bayesian} point out that Bayesian model averaging is an approach to helping arbitrate between competing predictors in a linear regression model. They average over all $2^6$  linear subset models excluding those that contain both variables \emph{Democratic proximity  advantage} and \emph{Democratic directional advantage}, (i.e., 48 models in total). Each subset regression is with the form
 $$ M_\gamma: y | X, \beta_0, \beta \sim N(  \beta_0 + X_\gamma \beta_\gamma, \sigma^2)  . $$
 Accounting for the different complexity, they used the hyper-$g$ prior  \citep{liang2012mixtures}.
 Let $\phi$ to be the inverse of the variance  $\phi=\frac{1}{\sigma^2}$.  The hyper-$g$ prior with a prespecified hyperparameter $\alpha$ is,
 \begin{align*}
p(\phi)&\propto \frac{1}{\phi}, \\
\beta | g, \phi, X &\sim \mbox{N} \Bigl(0, \frac{g}{\phi} (X^TX)^{-1}\Bigr),  \\
p (g | \alpha ) &=\frac{ \alpha-2}{ 2} (1+g) ^{-\alpha/2}, \ g>0 . 
 \end{align*}

The first two columns of Figure \ref{Senate} show the linear regression coefficients as estimated using least squares.   The remaining columns show the posterior mean and standard deviation of the regression coefficients using BMA, stacking  of predictive distributions,  and Pseudo-BMA+, respectively. Under all three averaging strategies, the coefficient of \emph{proximity advantage} is no longer statistically significantly negative, and the coefficient of \emph{directional advantage} is shrunk.  As fit to these data, stacking puts near-zero weights on all subset models containing \emph{proximity advantage}, whereas  Pseudo-BMA+ weighting always gives some weight to each model. In this example,  averaging subset models by stacking or Pseudo-BMA+ weighting  gives a way to deal with competing variables, which should be more reliable than BMA according to our previous argument.

\subsection{Predicting well-switching behavior in Bangladesh}
Many wells in Bangladesh and other South Asian countries are contaminated with natural arsenic. People whose wells have arsenic levels that exceed a certain threshold are encouraged to switch to nearby safe wells \citep[for background details, see Chapter 5 in][]{gelman2006data}. We are analyzing a dataset  including $3020$ respondents to find factors predictive of the well switching among all people with unsafe wells.  The outcome variable is 
$$ y_i= 
\begin{cases}
1& \text{ if household $i$ switched to a new well }\\
0& \text{ if household $i$ continued using its own well.}
\end{cases}$$
And we consider the following inputs:
\begin{itemize}
\item The distance (in meters) to the closest known safe well,
\item The arsenic level of the respondent's well,
\item Whether any member of the household is active in the community association,
\item The education level of the head of the household.
\end{itemize}

We start with what we call Model 1, a simple logistic regression with all variables above as well as a constant term:
\[ \begin{split}
y\sim &\mathrm{Bernoulli} (\theta)\\
\theta= & \mathrm{logit}^{-1}(\beta_0+\beta_1 dist +\beta_2 arsenic+\beta_3 assoc+ \beta_4 edu)
\end{split} \]
Model 2 contains the interaction between distance and arsenic level.
$$\theta= \mathrm{logit}^{-1}(\beta_0+\beta_1 dist +\beta_2 arsenic+\beta_3 assoc+ \beta_4 edu +\beta_5 dist\times arsenic  )$$
Furthermore, it makes sense to us a nonlinear model for logit switching probability as a function of distance and arsenic level. We can use spline to capture that.  Our Model 3 contains the  B-splines for distance and arsenic level with polynomial degree 2, 
$$\theta=  \mathrm{logit}^{-1}(\beta_0+\beta_1 dist +\beta_2 arsenic+\beta_3 assoc+ \beta_4 edu +\alpha_{dis} B_{dis} +\alpha_{ars} B_{ars})$$
where $B_{dis}$ is the  B-spline basis of distance with the form $\bigl(B_{dis, 1} (dist), \dots, B_{dis, q} (dist) \bigr)$  and $\alpha_{dis}, \alpha_{ars} $ are vectors. We also fix the number of knots to be 10 for both distance and arsenic level. Model 4 and 5 are the similar models with 3-degree and 5-degree B-splines, respectively.

\begin{figure}
 \includegraphics[width=\textwidth ] {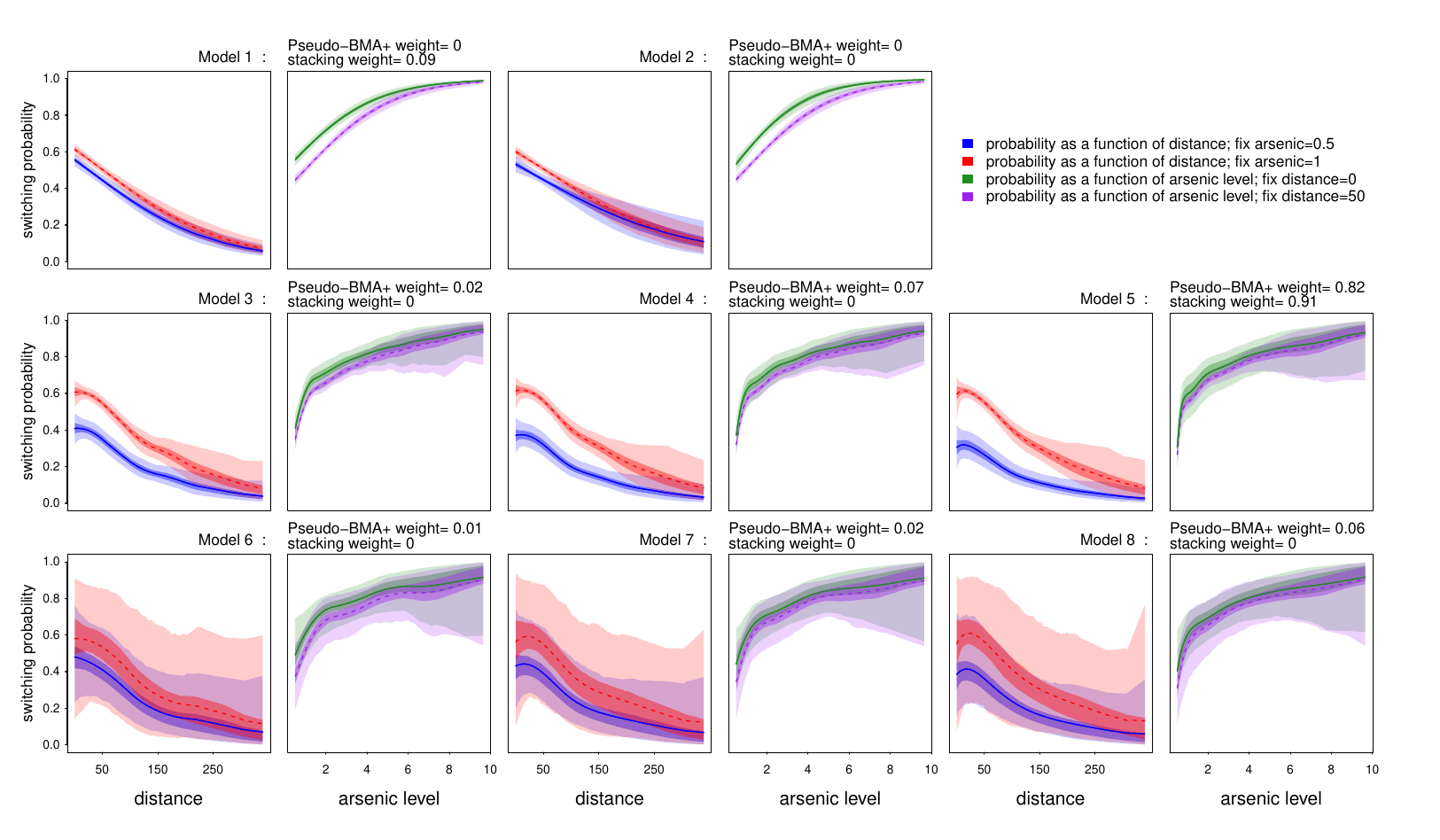}
\vspace{-.3in}
\caption{\em Posterior mean, 50\% confidence interval, and 95\% confidence interval  of the probability of switching from an unsafe well  in Models 1--8.  For each model, the switching probability is shown  as a function of (a) the distance to the nearest safe well or (b) the arsenic level of the existing well. For each plot, the other input variable is held constant at different  representative  values.  The model weights by stacking of predictive distributions and Pseudo-BMA+ are listed above each panel.} \label{well}
 \end{figure}       

Next, we can add a bivariate spline to capture a nonlinear interaction,
$$\theta=  \mathrm{logit}^{-1}(\beta_0+\beta_1 dist +\beta_2 arsenic+\beta_3 assoc+ \beta_4 edu   +\beta_5 dist\times arsenic+\alpha B_{dis, ars})$$
where $B_{dis, ars}$ is the bivariate spline basis with the degree to be $2\times 2, 3\times 3, 5\times 5$ in Model $6,7$ and $8$ respectively. 

Figure \ref{well} shows the inference results in all 8 models, which are summarized by the posterior mean, 50\% confidence interval and 95\% confidence interval  of the probability of switching from an unsafe well as a function of distance or arsenic level.  Any other variables such as {\tt assoc} and {\tt edu}  are fixed at their means.  It is not obvious to pick one best model. Spline models give a more flexible shape, but also introduce more variance for posterior estimation. 

\begin{figure}
 \includegraphics[width=\textwidth ] {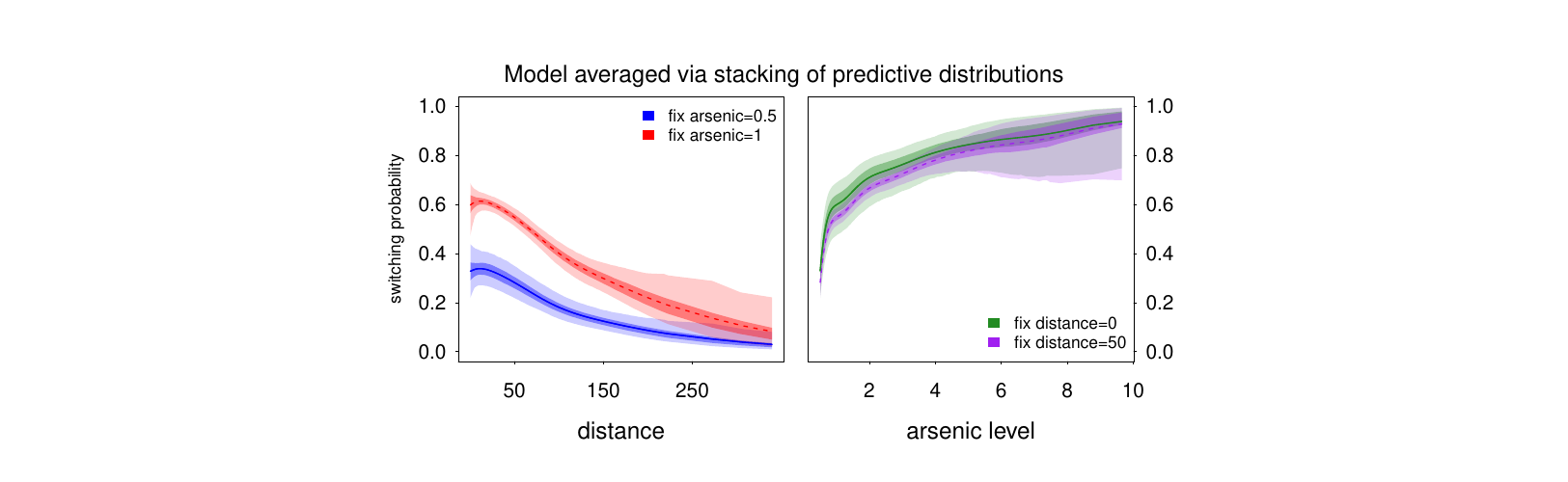}
\vspace{-.3in}
\caption{\em Posterior mean, 50\% confidence interval, and 95\% confidence interval  of the probability of switching from an unsafe well in the combined model via stacking of predictive distributions.  Pseudo-BMA+ weighting gives a similar result for the combination.} \label{well-combine}
 \end{figure}

Finally, we run stacking of predictive distributions and Pseudo-BMA+ to combine these 8 models. The calculated model weights are listed above each panel in Figure \ref{well}.  For both combination methods, Model 5 (univariate splines with 5th degree) accounts for the majority share.   It is also worth pointing out that Model 8 is the most complicated one, but both stacking and Pseudo-BMA+ avoid  overfitting by assigning a very small weight on it.

Figure \ref{well-combine} shows the posterior mean, 50\% confidence interval, and 95\% confidence interval  of the switching probability in the stacking-combined model.  Pseudo-BMA+ weighting gives a similar combination result for this example.  At first glance, the combination looks quite similar to Model 5, while it may not seem necessary to put an extra 0.09 weight on Model 1 in stacking combination since Model 1 is completely contained in Model 5 if setting $\alpha_{dis} =\alpha_{ars}=0$. However, Model 5 is not perfect since it predicts that the posterior mean of switching probability will decrease as a function of distance to the nearest safe well, for very small distances.  In fact, without further control, it is not surprising to find boundary fluctuation as a main drawback for higher order splines. Fortunately,  we notice this decrease trend  around the left boundary is a little bit flatter in the combined distribution since the combination contains part of straightforward logistic regression (in stacking weights) or lower order splines (in Pseudo-BMA+ weights).   In this example the sample size $n=3020$ is large, hence we  have reasons to believe stacking of predictive distributions gives the optimal combination.

\section{Discussion}

\subsection {Sparse structure and high dimensions}

\cite{yang2014minimax} propose to estimate a linear combination of point forecasts, 
$f= \sum_ {k=1}^{K} w_k  f_k$, using a  Dirichlet aggregation prior, 
$w \sim \mathrm{Dirichlet} \bigl(  \frac{\alpha}{ K^\gamma},  \dots   , \frac{\alpha}{ K^\gamma} \bigr)$, to pull toward structure, and estimating the weights $w_k$ using adaptive regression rather than cross-validation.  They show that the combination under this setting can achieve the minimax squared risk among all convex combinations,
$$ \sup_{f_1,\dots f_K \in F_0 }  \inf _{\hat f} \sup _{f_\lambda^* \in F_\Gamma } E || \hat f- f_\lambda^*   || ,  $$
where $F_0=(  f: ||f||_\infty \leq 1  ).$ 

Similar to our problem, when the dimension of model space is high, it can make sense to assign a strong prior can to the weights in estimation equation (\ref{stacking}) to improve the regularization, using a hierarchical prior to pull toward sparsity if that is desired.

\subsection{Constraints and regularity}
In  point estimation stacking,  the simplex constraint is  the most widely used regularization so as to overcome potential problems with multicollinearity.  \cite{clarke2003} suggests relaxing the constraint to make it more flexible.

When combining distributions,  there is no need to worry about multicollinearity except in degenerate cases. But in order to guarantee a meaningful posterior predictive density, the simplex constraint becomes natural, which is satisfied automatically in BMA and Pseudo-BMA weighting.  As mentioned in the previous section, stronger priors can be added. 
 
Another assumption is that the separate posterior distributions are combined linearly and with weights that are positive and sum to 1.
There could be gains from going beyond convex linear combinations.  For instance, in the subset regression example when each individual model is a univariate regression,  the true model distribution is a convolution instead of a mixture of each  possible models distribution.  Both of them lead to the additive model in the point estimation, so stacking of the means is always valid, while stacking of predictive distributions  is not possible to recover the  true model in the convolution case.
 
 Our explanation is that when the model list is large, the convex span should be large enough to approximate the true model. And this is the reason why we prefer adding stronger priors to make the estimation of weights stable in high dimensions.

\subsection{General recommendations}
The methods discussed in this paper are all based on the idea of fitting models separately and then combining the estimated predictive distributions.  This approach is limited in that it does not pool information between the different model fits:  as such, it is only ideal when the $K$ different models being fit have nothing in common.  But in that case we would prefer to fit a larger super-model that includes the separate models as special cases, perhaps using an informative prior distribution to ensure stability in inferences.

That said, in practice it is common for different sorts of models to be set up without any easy way to combine them, and in such cases it is necessary from a Bayesian perspective to somehow aggregate their predictive distributions.  The often-recommended approach of Bayesian model averaging can fail catastrophically in that the required Bayes factors can depend entirely on arbitrary specifications of noninformative prior distributions.  Stacking  is a more promising general method in that it is directly focused on performance of the combined predictive distribution.  Based on our theory, simulations, and examples, we recommend stacking  (of predictive distributions) for the task of combining separately-fit Bayesian posterior predictive distributions.  As an alternative,  Pseudo-BMA+ is computationally cheaper and can serve as an initial guess for stacking. The computations can be done in R and Stan, and the optimization required to compute the weights connects directly to the predictive task.

\section*{Appendix A.  Implementation in Stan and R}
The $n\times K$ matrix of cross-validated log likelihood values,  $\{p(y_i | y_{-i} , M_k)\}_{i=1,\dots, n, k=1,\dots, K}$,  can be computed from the generated quantities block in a Stan program, following the approach of \cite{practicalPSIS}.  For the example in Section \ref{reg},  the $k$-th model is a linear regression with the  $k$-th covariates. We put the corresponding Stan code in the file  \texttt{regression.stan}:
\begin{verbatim}
data {
  int n;
  int p;         
  vector[n] y;
  matrix[n, p] X;
}
parameters {
  vector[p] beta;
  real<lower=0> sigma;
}
transformed parameters {
  vector[n] theta;
  theta = X * beta;
}
model {
  y ~ normal(theta, sigma);
  beta ~ normal(0, 10);
  sigma ~ gamma(0.1, 0.1);
}
generated quantities {
  vector[n] log_lik;
  for (i in 1:n)
    log_lik[i] = normal_lpdf(y[i] | theta[i], sigma);
}
\end{verbatim}
In R we can simulate the likelihood matrices from all $K$ models and save them as a list:
\begin{verbatim}
library("rstan")
log_lik_list <- list()
for (k in 1:K){
  # Fit the k-th model with Stan
  fit <- stan("regression.stan", data=list(y=y, X=X[,k], n=length(y), p=1))    
  log_lik_list[[k]] <- extract(fit)[["log_lik"]]
}
\end{verbatim}
The function \texttt{model\_weights()} in the \texttt{loo} package\footnote{ The package can be downloaded from  https://github.com/stan-dev/loo/tree/yuling-stacking} in R can give model combination weights according to stacking of predictive distributions, Pseudo-BMA  and Pseudo-BMA+ weighting. We can choose whether to use the Bayesian bootstrap to make further regularization for Pseudo-BMA, if computation time is not a concern.
\begin{verbatim}
model_weights_1 <- model_weights(log_lik_list, method="stacking")
model_weights_2 <- model_weights(log_lik_list, method="pseudobma", BB=TRUE)
\end{verbatim}
in one simulation with six models to combine, the output gives us the computed weights under each approach:
\begin{verbatim}
The stacking weights are:
[1,] "Model 1" "Model 2" "Model 3" "Model 4" "Model 5" "Model 6"
[2,] "0.25"    "0.06"    "0.09"    "0.25"    "0.35"    "0.00"  

The Pseudo-BMA+ weights using Bayesian Bootstrap are:
[1,] "Model 1" "Model 2" "Model 3" "Model 4" "Model 5" "Model 6"
[2,] "0.28"    "0.05"    "0.08"    "0.30"    "0.28"    "0.00"    
\end{verbatim}
For reasons discussed in the paper, we generally recommend stacking for combining separate Bayesian predictive distributions.

\vskip 0.2in
\bibliographystyle{ba}
\bibliography{combine}

 \begin{acknowledgement}
 We thank the U.S. National Science Foundation, Institute for Education Sciences, Office of Naval Research, and Defense Advanced Research Projects Administration for partial support of this work.  We also thank the Editor, Associate Editor, and two anonymous referees for their valuable comments that helped strengthen this manuscript.
 \end{acknowledgement}

\end{document}